\documentstyle[preprint,eqsecnum,aps,epsf]{revtex}	

\newif\iftightenlines\tightenlinesfalse
\tightenlines\tightenlinestrue

\begin{document}
%
\def\eslt{E\llap/_T}
\def\esl{E\llap/}
\def\msl{m\llap/}
\def\to{\rightarrow}
\def\te{\tilde e}
\def\tmu{\tilde\mu}
\def\ttau{\tilde\tau}
\def\tl{\tilde\ell}
\def\ttau{\tilde \tau}
\def\tg{\tilde g}
\def\tnu{\tilde\nu}
\def\tell{\tilde\ell}
\def\tq{\tilde q}
\def\tb{\tilde b}
\def\tst{\tilde t}
\def\tt{\tilde t}
\def\tw{\widetilde W}
\def\tz{\widetilde Z}

\hyphenation{mssm}
%
\preprint{\vbox{\baselineskip=14pt%
   \rightline{FSU-HEP-950501}\break
   \rightline{UH-511-829-95}
}}
\title{PROSPECTS FOR SUPERSYMMETRY AT LEP2}
\author{Howard Baer$^1$, Michal Brhlik$^1$, Ray Munroe$^1$ and Xerxes Tata$^2$}
\address{
$^1$Department of Physics,
Florida State University,
Tallahassee, FL 32306 USA
}
\address{
$^2$Department of Physics and Astronomy,
University of Hawaii,
Honolulu, HI 96822 USA
}
\date{\today}
\maketitle
\begin{abstract}

Working within the framework of the minimal supergravity model with
gauge coupling unification and radiative electroweak symmetry breaking
(SUGRA), we map out regions of parameter space explorable by
experiments at LEP2, for center of mass energy options of
$\sqrt{s}=150,\ 175$, $190$ and 205  GeV.
We compute signals from all accessible $2 \rightarrow 2$ SUSY pair
production processes using the ISAJET simulation program, and devise cuts that
enhance the signal relative to Standard Model backgrounds, and which also
serve to differentiate various supersymmetric processes from one another. We
delineate regions of SUGRA parameter space where production of neutralino
pairs, chargino pairs, slepton pairs and the production of the light Higgs
scalar of SUSY is detectable above Standard Model backgrounds and
distinguishable from other SUSY processes.
In addition, we find small regions of SUGRA parameter space where $\te\te$,
$\tz_2\tz_2$ and $\tnu_L\tnu_L$ production yields
spectacular events with up to four
isolated leptons.
The combined regions of parameter space explorable by LEP2 are
compared with the reach of Tevatron Main Injector era experiments.
Finally, we comment on how the reach via the neutralino pair channel is
altered
when the radiative electroweak symmetry breaking constraint is relaxed.
\end{abstract}

\medskip
\pacs{PACS numbers: 14.80.Ly, 13.85.Qk, 11.30.Pb}



\section{Introduction}

The CERN $e^+e^-$ collider LEP, currently running with total center-of-mass
energy around the $Z$ pole, is expected to undergo an energy upgrade in the
near future, to become LEP2. The machine energy will ultimately exceed the WW
production threshold so that experiments at LEP2 will directly probe the
form of the
ZWW and the $\gamma W W$ interactions \cite{VV}. The higher energy and the
clean
experimental environment of LEP2 will also allow direct searches
for new particles,
including the Higgs boson, the expected relic of the spontaneous breaking of
the electroweak gauge group.
Another important goal of LEP2 experiments
is the direct search for new particles that occur in
various extensions of the Standard Model (SM),
the most promising of which is low energy supersymmetry\cite{MSSM,DPF}.
Already, the four LEP
experiments have placed relatively model independent bounds on the masses of
various sparticles and Higgs bosons\cite{LEP}. To be specific\cite{PDB},
\begin{eqnarray*}
&m&_{\tw_1}>45\ {\rm GeV},\\
&m&_{\tell}>45\ {\rm GeV},\ (\tell =\te ,\tmu , \ttau ) \\
&m&_{\tq}>45\ {\rm GeV},\\
&m&_{\tnu}>41.8\ {\rm GeV},\ ({\rm three\ degenerate\ flavors}) \\
&m&_{H_{\ell}}>44\ {\rm GeV},\ ({\rm for\ \tan\beta >1}),
\end{eqnarray*}
where $\tw_1$ is the lightest chargino and $H_{\ell}$ is the lightest
neutral scalar in the Higgs sector.
The above sparticle mass limits are mainly limited by the beam energy.
Hence, considerable improvement is expected at LEP2.
In addition, the CDF and D0 collaborations, from a non-observation
of any excess of $\eslt$ events at the Fermilab Tevatron $p\bar p$
collider, now require\cite{CDFDZERO}
\begin{eqnarray*}
m_{\tg}&>&150\ {\rm GeV},\\
m_{\tq}&>&150\ {\rm GeV},\ ({\rm if\ m_{\tg}\alt 400\ GeV}).
\end{eqnarray*}
The Tevatron bounds have been obtained within the framework of
the Minimal Supersymmetric Model (MSSM) and are somewhat
sensitive to the assumed unification of gaugino masses, but depend only weakly
on other SUSY parameters.

Many previous LEP analyses\cite{OLD1,OLD} (including the experimental
ones) have been performed within the
framework of the supergravity-inspired MSSM. The weak
scale sparticle masses are assumed to originate from unification scale
common soft breaking terms $m_0$ (for scalar sparticles) and $m_{1/2}$
(for gaugino masses). Thus the first five flavors of squarks
are assumed to be approximately degenerate, as are the sleptons.
The soft-breaking trilinear coupling $A_t$ mainly affects the mass
and the phenomenology
of top squarks, and is
neglected for most purposes. The ratio $\tan\beta$ of the two Higgs
field vacuum expectation values,
the SUSY-conserving superpotential Higgs mass $\mu$, and finally,
the pseudoscalar Higgs boson mass $m_{H_p}$ are taken to be free
parameters. These analyses
generally focus upon the production of just one sparticle
species at a time, although 31 new particles are predicted, and it is
possible to have several closely spaced thresholds.

Recently, several groups\cite{COLL,LOPEZ,BCMPT,TEVST,MRENNA} have studied SUSY
phenomenology at colliders
within the framework of the highly constrained minimal
supergravity (SUGRA) grand unified model,
with gauge coupling unification and radiative electroweak symmetry breaking.
SUGRA models should be regarded as effective theories with Lagrangian
parameters renormalized at
an ultra-high scale $M_X \sim M_{GUT}-M_{Planck}$, and valid only
below this scale.
The corresponding weak scale sparticle coupling and masses are
then calculated by evolving 26 renormalization group equations \cite{RGE} from
the unification scale to the weak scale. An elegant
by-product \cite{RAD} of this mechanism
is that one of the Higgs boson mass squared terms is driven negative,
resulting in a breakdown of electroweak symmetry. This model is completely
specified by four \cite{NILLES}
SUSY parameters (in addition to SM masses and couplings).
A hybrid set consisting of the common mass $m_0$ ($m_{1/2}$) for all scalars
(gauginos), a common SUSY-breaking trilinear coupling $A_0$ all specified
at the scale $M_X$ together with $\tan\beta$ proves to be a convenient choice.
These parameters fix the masses and couplings of all sparticles. In
particular, $m_{H_p}$ and the magnitude (but not the sign) of $\mu$ are fixed.
In other words, various assumptions about the symmetries
of interactions at the scale $M_X$ that have been built into the SUGRA
framework restrict the model parameters to
a subset of the SUGRA-inspired MSSM parameter space referred to earlier.

The SUGRA framework (and also a SUGRA-inspired MSSM framework
without radiative elecroweak symmetry breaking)
has been incorporated into the event
generator program ISAJET 7.13\cite{BCMPT,ISAJET}.
All lowest order $2\to 2$ sparticle and Higgs boson
production mechanisms have been incorporated into ISAJET. These include
the following processes (neglecting bars over anti-particles):
\begin{eqnarray*}
e^+e^-&\to & \tq_L\tq_L,\ \tq_R\tq_R ,\\
e^+e^-&\to & \tl_L\tl_L,\ \tl_R\tl_R,\ \te_L\te_R ,\\
e^+e^-&\to & \tnu_{\ell}\tnu_{\ell},\\
e^+e^-&\to & \tw_1\tw_1,\ \tw_2\tw_2,\ \tw_1\tw_2 ,\\
e^+e^-&\to & \tz_i\tz_j,\ (i,j=1-4),\\
e^+e^-&\to & Z H_{\ell},\ Z H_h,\ H_p H_{\ell},\ H_p H_h,\ H^+ H^-.
\end{eqnarray*}
In the above, $\ell =e,\ \mu$ or $\tau$. All squarks (and also all sleptons
other than staus) are taken to be
$L$ or $R$ eigenstates, except the stops, for which $\tst_1\tst_1$,
$\tst_1\tst_2$ and $\tst_2\tst_2$ (here, $\tst_{1,2}$ being the lighter/heavier
of the top squark mass eigenstates) production is included.
Given a point in SUGRA space, and a collider energy, ISAJET generates all
allowed production processes, according to their relative cross sections.
The produced sparticles or Higgs bosons are then decayed into all
kinematically accessible
channels, with branching fractions calculated within ISAJET.
The sparticle decay cascade terminates with the
lightest SUSY particle (LSP), taken to be the lightest neutralino ($\tz_1$).
Final state QCD radiation is included, as well as particle hadronization.
ISAJET currently
neglects spin correlations, sparticle decay
matrix elements, and also, initial state photon radiation.
In the above reactions, spin correlation effects are only important
for chargino
and neutralino pair production,
while decay matrix elements are only important
for 3-body sparticle decays.

The purpose of this paper is three-fold.
\begin{enumerate}
\item We examine SUSY signals in the highly restricted SUGRA framework.
We note that frequently
one must consider not just a single SUSY production mechanism, but rather
one must often consider simultaneously production of several different
sparticles, since their masses are expected to be correlated.
For the purposes of sparticle detection, this
means that not only should signals be observable above SM backgrounds, but
also that two or more signals have to be untangled from one another if
they happen to occur simultaneously.

\item We examine how the regions of SUGRA parameter space for sparticle or
Higgs boson detection alter as a function of the machine energy and luminosity.
We are
motivated by the possibility that it may be feasible to increase the energy
of LEP2 from its starting value of about 140-150~GeV.
The options considered are
\begin{itemize}
\item $\sqrt{s}=150$ GeV, $\int {\cal L}dt=500$ pb$^{-1}$,
\item $\sqrt{s}=175$ GeV, $\int {\cal L}dt=500$ pb$^{-1}$,
\item $\sqrt{s}=190$ GeV, $\int {\cal L}dt=300$ pb$^{-1}$,
\item $\sqrt{s}=205$ GeV, $\int {\cal L}dt=300$ pb$^{-1}$.
\end{itemize}
The first of these cases is of special interest\cite{OLD1,OLD} because, below
the $WW$
threshold, SM backgrounds are frequently tiny.

\item There are regions of parameter space where the only visible sparticle
production could come from $\tz_1\tz_2$ production. Within the MSSM, the
rate for this reaction can be very small if the neutralinos are mainly
gaugino-like and sleptons heavy, which is probably why this reaction has
not been studied in as much detail as chargino or slepton pair production
in the earlier literature. We examine the prospects
of identifying signals from this reaction over SM backgrounds
and further, of discriminating $\tz_1\tz_2$ production from other
SUSY and Higgs production processes.

\end{enumerate}

As an illustration of (1) above, we show in Fig.~1 total
sparticle production cross sections at $\sqrt{s}=175$ GeV versus the
unification scale gaugino mass $m_{1/2}$. We take $A_0=0$, $\tan\beta =2$
and $\mu <0$. In Fig. 1{\it a}, we take $\xi={m_0\over m_{1/2}}=0$, and begin
the lower limit of our plot from $m_{1/2}\sim 90$ GeV, below which the
sneutrino mass violates the above LEP bounds (Eq. (1)). For $m_{1/2}<140$ GeV,
pair production of $L$- and $R$- selectrons is dominant, although
$\sigma (e^+e^-\to ZH_{\ell})\sim 1000$ fb. In addition, smuon and stau
pair production is taking place at $\sigma\sim 500$ fb.
Chargino pair production is kinematically forbidden but
$\sigma (e^+e^-\to\tz_2\tz_1 )\sim$ 200-400 fb, when $m_{1/2}$ is small.
In this case, the neutralino pair signals may be difficult to extract
from a background which includes {\it other} SUSY and Higgs boson processes.
In Fig. 1{\it b}, we take $\xi=1$. In this case, sleptons are too heavy to
be produced and the dominant new-particle
cross section comes from $ZH_{\ell}$ production, followed by $\tz_2\tz_1$
and $\tw_1\overline{\tw_1}$ production (which just becomes accessible)
when $m_{1/2}$ is small.
Finally, in Fig. 1{\it c} we take $\xi=4$. Now, because
smaller values of $m_{1/2}$ are not excluded by LEP experiments,
$\tw_1\overline{\tw_1}$
production is dominant out to $m_{1/2}=90$ GeV, followed by $ZH_{\ell}$
production. The production of $\tz_2\tz_1$ events occurs at a very low rate,
and would be difficult to separate from the $\tw_1\overline{\tw_1}$ pair
signals, as well as the SM $WW$ background.

The remainder of this paper is organized as follows. In Sec. II, III and IV,
we describe our analyses for the extraction of various signals at centre of
mass energies of 150, 175 and 190-205~GeV, respectively. Because
$WW$ and $ZZ$ ($ZZ$) production is kinematically inaccessible in the first
(second) case, SM physics backgrounds, (and hence the cuts we choose to
extract the signal) differ in the three cases. The reader who is not interested
in the details of the analysis need focus only on the results presented in
the figures and the accompanying discussion, but can skip over the details
of the selection criteria detailed therein. In Section V, we discuss
additional signals such
as $4\ell$ production
via {\it e.g.} $\tnu_{\ell}\overline{\tnu_{\ell}}$, $\te\overline{\te}$
or $\tz_2\tz_2$
production, which don't necessarily extend the parameter space reach, but do
yield exotic, gold-plated signatures for sparticle production reactions that
are usually neglected in the literature. In Sec. VI we compare
the reach of the various
LEP2 upgrade options amongst themeselves, and with
the capabilities of the Fermilab Tevatron Main Injector upgrade.
In Sec. VII, we discuss
how neutralino signals (in particular) are altered if
the constraint from radiative electroweak symmetry breaking is relaxed. The
latter allows $|\mu |$ to become a free parameter, so that the lighter $\tz_i$
can have large higgsino components.
We focus mainly on neutralino production because
pair production of
charged sparticles is less model dependent, and their signals
have also been
extensively\cite{OLD1,OLD} studied within the SUGRA-inspired MSSM framework.
We conclude with a summary of our results in Sec. VIII.

\section{Sparticle signals at $E_{CM}=150$ GeV.}

The first test run of LEP2 is expected to begin in late 1995, with
collider energy of $\sqrt{s}\sim 140-150$ GeV. A major feature of a collider
run at this energy is that it is still below threshold for $WW$ production,
and SM backgrounds to signals from the production of new, heavy particles
are small. Indeed,
if hints of new physics signal are seen, then collection of substantial
integrated luminosity below $WW$ threshold may be desirable \cite{OLD1,OLD}.
Furthermore, in this energy range, one does not expect to produce the light
Higgs scalar $H_\ell$. In the minimal SUGRA model, where $|\mu |$ (derived from
radiative electroweak symmetry breaking) is typically large, $H^{\pm}$, $H_h$
and $H_{p}$ are very heavy, whereas
the light Higgs scalar
is expected to be nearly indistinguishable from a SM Higgs scalar. Hence,
except in some corners of parameter space,
LEP limits of $m_{H_{SM}}\agt 60$ GeV apply as well to $H_\ell$, so that a
collider energy of $\sqrt{s}>m_Z + 60\ \sim 150$ GeV will be needed to probe
new territory in the Higgs sector. The reactions to focus on at
$\sqrt{s}\sim 150$ GeV are then {\it i}) $\tl_i\overline{\tl_i}$ production
(where $\tl =e$, $\mu$ or $\tau$, and $i=L$ or $R$),
{\it ii}) $\tw_1\overline{\tw_1}$
production, and {\it iii}) $\tz_1\tz_2$ production.

We do not consider
squark signals in this paper, since squarks light enough to be accessible at
LEP2 are already excluded by hadron collider data. A possible exception
is the light $\tt_1$ for which the hadron collider limits are not applicable.
The best limit on $m_{\tt_1}$ come from LEP experiments,
so that LEP2 should be able to probe beyond the current bounds. With a data
sample of about 100~$pb^{-1}$ the Tevatron experiments will also be able
to probe\cite{SENDER} $\tt_1$ masses up to 80-100~GeV.
For this reason, and because
$m_{\tt_1}$ is rarely lighter than 100 GeV in SUGRA
parameter space, we do not consider top squark signals any further in this
paper.

{\it 2.1 Selectrons:}
Although potentially any of the
slepton pair reactions ({\it e.g.}
$\te_R\overline{\te_R}$, $\te_R\overline{\te_L}$, $\te_L\overline{\te_R}$,
$\te_L\overline{\te_L}$, $\tmu_R\overline{\tmu_R}$, $\tmu_L\overline{\tmu_L}$,
$\ttau_1\overline{\ttau_1}$, $\ttau_2\overline{\ttau_2}$) can occur at LEP,
we focus only on the selectron pair
production reactions. Unlike smuon or stau production which occurs only
via $s$-channel $\gamma$ and $Z$ exchanges, selectron pair production
can also occur via the exchange of neutralinos in the $t$-channel. Because
the left- (right-) slepton masses are expected to be independent of flavour
(except for negligible effects from the differences in Yukawa interactions),
the additional $t$-exchange contributions generally result in larger cross
sections for selectron pair production than for the production of smuon or
stau pairs\cite{OLD1,OLD}.
As a result, smuons and staus can usually be detected in a subset of the
parameter space where selectrons are observable, although for very large
values of $\tan\beta$ where the stau mixing induced by tau Yukawa
interactions becomes important, it is possible that $\ttau_1\overline{\ttau_1}$
production is the only accessible slepton production process.

Selectron pair production usually results in a very clean event
containing an acollinear $e^+e^-$ pair plus missing energy. Below
the $WW$ threshold, the main backgrounds come from {\it i})~$\tau^+\tau^-$
production followed by the leptonic decays of the $\tau$'s,
{\it ii})~$e^+e^-\gamma$ production, where the photon is lost down the beam
pipe,
and {\it iii}) $e^+ e^- e^+ e^-$ production via two photon reactions.
It has been shown that \cite{OLD} requiring
$\eslt > \sqrt{s}\sin\theta_{min}/(1+\sin\theta_{min})$, where
$\theta_{min}$ is the minimum angle above which leptons and photons can be
efficiently detected, very effectively
eliminates background from processes ({\it ii}) and ({\it iii}).
At LEP2 energies, the leptons from $\tau^+\tau^-$ production are essentially
back-to-back in the transverse plane.
To quantify the size of the signal and the $\tau$ pair background,
we generate selectron pair events as well as $\tau^+\tau^-$ background using
ISAJET. We require,

\begin{eqnarray}
& E_\ell & > 3\ {\rm GeV},\ \ \ |\eta_{\ell}|<2.5, \\
&\eslt &> 7.5\ {\rm GeV},\\
&\cos &\phi(\ell^+\ell^- ) > -0.9.
\end{eqnarray}

After these cuts, we are left with no significant SM background at
$\sqrt{s}=150$~GeV. We then (conservatively) require 10 signal events
to claim discovery.

The regions of selectron observability in SUGRA parameter
space are plotted in the $m_0\ vs.\ m_{1/2}$ plane in Fig.~2, where we
take $A_0=0$ and $m_t=170$ GeV. In Fig.~2 frame {\it a}), we show results for
$\tan\beta =2$ and $\mu <0$, while in {\it b}) we take the same $\tan\beta =2$
but require $\mu >0$. Finally, in {\it c}), we take $\tan\beta =10$ with
$\mu <0$, and in {\it d}) we take $\tan\beta =10$ with $\mu >0$.
Regions excluded by theoretical constraints such as lack of
appropriate electroweak symmetry breaking, or where the LSP is not $\tz_1$, are
enclosed by solid contours, and labelled TH. Similarly, regions excluded
by various LEP constraints
($m_{\tw_1}<47$ GeV \cite{FN1}, $m_{\tnu}<43$ GeV, $m_{H_\ell}<60$ GeV)
and Tevatron constraints from multi-jets$+\eslt$ searches are denoted by EX.
The regions of selectron observability are denoted by dashed contours
in the low $m_0$ region. Generally, selectrons are observable over most of
the region where their production is kinematically allowed. An exception to
this, however, occurs around $m_0\sim 0$, and $m_{1/2}\sim 150$ GeV
in case {\it a}) where the contour turns over. In this
region, the mass difference $m_{\tell_R}-m_{\tz_1}$
becomes so small that there is not enough visible energy from the selectron
decays to yield an observable signal.

{\it 2.2 Charginos:}
The chargino pair production cross section is typically in a few $pb$ range
when chargino pair production is kinematically allowed, although
it may be significantly suppressed when $m_{\tnu}\sim {\sqrt{s}\over 2}$.
Chargino pair signals occur in the multi-jet $+\eslt$ channel, the mixed
$\ell +$jet(s)$+\eslt$ channel, and the $\ell\overline{\ell'}+\eslt$ channel,
any
of which might be readily
observable when LEP2 is operating below $WW$ threshold.
The cuts below have been suggested\cite{GRIVAZ} for a chargino search when
$\tw_1\overline{\tw_1}\to \ell\nu\tz_1\ +\ q\bar{q'}\tz_1$. Although these
cuts are optimized for a chargino search above $WW$ threshold, they
generally allow a search for charginos up to threshold even for
$\sqrt{s}\sim 150$ GeV. Hence, we require,
\begin{eqnarray}
&\# & {\rm charged\ particles} > 5,\\
&\eslt &> 10\ {\rm GeV},\\
&{\rm isolated}&\ e\ {\rm or}\ \mu\ {\rm with}\ E_\ell > 5\ {\rm GeV},\\
&{\rm missing}& \ {\rm mass}\ > 63\ {\rm GeV},\\
&{\rm mass}& \ {\rm of\ the\ hadronic\ system} < 45\ {\rm GeV},\\
&m(\ell\nu )&  < 70\ {\rm GeV}.
\end{eqnarray}

The ten event regions where the signal is nominally taken to be observable
are below the dot-dashed contours
in Fig.~2,
and roughly follow the contour of constant $m_{1/2}$.
The notable exceptions occur
in Fig. 2{\it c} and 2{\it d}, where the dot-dashed contour turns down at
$m_0\sim 50$ GeV.
Below this value of $m_0$, the decay mode $\tw_1\to\tnu_{\ell L}\ell$ turns
on, so only the $\tw_1\overline{\tw_1}\to\ell\overline{\ell'}+\eslt$ mode
occurs.
If $m_{\tnu_{\ell L}}\sim m_{\tw_1}$, then the final state leptons are
very soft, and may be difficult to detect. For even lower values of $m_0$,
the sneutrinos become even lighter, and the purely leptonic channel from
chargino pair production can fill in {\it part} of the gap
between the selectron
contour and the dashed-dotted chargino pair contour.

{\it 2.3 Neutralinos:}
Neutralino pair production occurs via $s$-channel $Z$ exchange and $t$- and
$u$-channel
$L$- and $R$- selectron exchange graphs.
In minimal SUGRA models, where $m_{\tz_1}\sim {1\over 2} m_{\tz_2}$ and
$m_{\tz_2}\sim m_{\tw_1}$, there exists a region of parameter
space where $\tz_1\tz_2$ is kinematically accessible but
$\tw_1\overline{\tw_1}$ production is forbidden.
Moreover, the two lightest neutralinos
are mainly gaugino-like, with small coupling to the $Z$ boson,
resulting in a suppressed contribution from the $s$-channel graph.
In addition, $t$-channel $\tz_1\tz_2$ production is suppressed when
$m_{\te_i}$ is heavy, but can be significant if $m_{\te_i}$
is light ($m_0$ small). Once produced, the $\tz_2$ can decay via real or
virtual $Z$, $H_i$, $\tl_i$, $\tnu$ or $\tq_i$. The most promising signatures
include $\tz_1\tz_2\to\ell\overline{\ell}+\eslt$ and $\tz_1\tz_2\to
q\overline{q}+\eslt$.
We have evaluated $e^+e^-\to\tz_1\tz_2$ along with decays without
spin correlations (using ISAJET), and with spin correlations
(using HELAS\cite{HELAS}), and find little difference between the final
signal rates. We attribute this to the fact that for SUGRA parameter space
regions where the signal is observable $\tz_1\tz_2$
production is dominated by slepton exchange, while decays are dominated
by squark, slepton and sneutrino exchange-- all spin-0 particles.

To search for $e^+e^-\to\tz_1\tz_2\to\ell\overline{\ell}+\eslt$ events, we
require
\begin{eqnarray}
& E_\ell &> 3\ {\rm GeV},\ \ \ |\eta_{\ell}|<2.5, \\
& \eslt &> 7.5\ {\rm GeV},\\
& \esl &> 89\ {\rm GeV},\\
& m&(\ell\overline{\ell}) < 55\ {\rm GeV},\\
&\phi &(\ell_1 ,\ell_2 ) < 172^o.
\end{eqnarray}
These allow one to see the dilepton signal above the SM $\tau\overline{\tau}$
background, and also above the dilepton level expected from
$\tw_1\overline{\tw_1}$ production. The resulting
ten event signal level is plotted in Fig. 2{\it a}-2{\it d} as the dotted
contour. In Fig. 2{\it a}, we see that $\tz_1\tz_2\to\ell\overline{\ell}+\eslt$
is visible mainly for $80\alt m_0 \alt 210$ GeV, for values of $m_{1/2}$
ranging up to $\sim 100$ GeV-- well beyond the reach for chargino pairs.
For smaller values of $m_0$, the signal is not observable because $\tz_2$
dominantly decays invisibly to $\nu\tnu$. In Fig. 2{\it b}, the
$\tz_1\tz_2\to\ell\overline{\ell}+\eslt$ signal should be
detectable for small
values of $m_0$ as well because here, $\tz_2\to\ell\tell_R$
decays dominate.
Note the small diagonal gap between the
two disjoint dilepton regions: here the $\tz_2\to\ell\tell_R$ decays
dominate, but are just barely open, and result in one of the signal
leptons being too soft to be observable. In Fig. 2{\it c},
there is again a small region of observability for
$\tz_1\tz_2\to\ell\overline{\ell}+\eslt$, but here it is limited to
$60 <m_0<100$ GeV, and does not give much additional region of observability
beyond the $\tw_1\overline{\tw_1}$ observability region. Finally, in
Fig. 2{\it d}, there exist three disjoint regions where
$\tz_1\tz_2\to\ell\overline{\ell}+\eslt$ is visible.

To search for $\tz_1\tz_2\to$ jets$+\eslt$, we first coalesce hadronic clusters
within a cone of $\Delta R<0.5$, and label as a jet if $E_j>5$ GeV and
$|\eta_j |<2.5$. In addition, we require
\begin{eqnarray}
&\eslt &> 7.5\ {\rm GeV},\\
&m&_{jet}>5\ {\rm GeV}\ (\rm mono-jet\ events),\\
&\phi &(j_1 ,j_2 ) < 172^o\ ({\rm di-jet\ events}).
\end{eqnarray}
We veto events with identified $e$ or $\mu$ in them.
The main SM background comes from $\tau\overline{\tau}$ production (recall that
$WW$ and $ZH_{\ell}$ production is inaccessible), and is
essentially eliminated by the latter two cuts.
However, in regions where $\tw_1\overline{\tw_1}$ is open, there is a large
background from $\tw_1\overline{\tw_1}\to$ 1 or 2 jets$+\eslt$, due to
double hadronic chargino decays where jets are soft or coalesce, and due
to single hadronic chargino decays, where the other chargino decays to a
hadronic $\tau$ or a soft or missing lepton. The latter supersymmetric
background makes $\tz_1\tz_2\to$ jets$+\eslt$ very difficult to distinguish
as an independent production and decay mechanism. In Fig. 2{\it a}-{\it d},
we show the region for hadronic $\tz_1\tz_2$ detection as short dashed
contours.
In Fig. 2{\it a}, $\tz_1\tz_2\to$ jets$+\eslt$ is observable
in a subset of the region where dileptons from $\tz_1\tz_2$ are visible.
In Fig. 2{\it b}, $\tz_1\tz_2\to$ jets$+\eslt$ is essentially not
visible at all, while in Figs. 2{\it c} and 2{\it d},
$\tz_1\tz_2\to$ jets$+\eslt$ is observable
beyond the $\tw_1\overline{\tw_1}$ region in a small but not
insignificant slice of parameter space.

\section{Sparticle production at $E_{CM}=175$ GeV.}

LEP2, running at $\sqrt{s}=175$ GeV, will be above threshold for $WW$
production, for which the total cross section is $\sigma (WW)\sim 17.3$ pb.
An extended run to gather $\int {\cal L}dt\sim 500$ pb$^{-1}$ of integrated
luminosity is expected to occur, to measure the $W$ mass and triple
vector boson coupling. Furthermore, at this energy, LEP2 will be
sensitive to higher
ranges of Higgs boson masses, up to $m_{H_{SM}}\sim 80$ GeV.

{\it 3.1 Selectrons:}
Selectron pair production has an
irreducible background now from $WW\to e^+\nu_e e^-\overline{\nu_e}$, which
occurs
at the 212 fb level. To reduce $WW$ background, additional cuts are needed
beyond those of (2.1)-(2.3):
\begin{eqnarray}
&3<& E_{\ell}<46\ {\rm GeV},\ |\eta_{\ell}|<2.5, \\
&\eslt &> 9\ {\rm GeV},\\
&\cos &\phi (\ell^+\ell^- )>-0.9, \\
&\pm &\cos\theta_{\ell^{\pm}}>0,
\end{eqnarray}
where $\theta_{\ell}$ is the angle between the beam and the detected lepton.
The $\cos\theta$ cut is applied only to the more energetic of the two detected
leptons.

To evaluate the background from $WW$ production, we have used the HELAS
package to calculate the spin-correlated Feynman diagrams for various
final state configurations (thus, the $WW$ and $ZZ$ backgrounds are evaluated
at the parton level). After the above cuts, we find backgrounds of
\begin{eqnarray*}
&WW&\to e^+ e^- :\ 11.5\ {\rm fb}, \\
&WW&\to \tau^\pm e^\mp \to e^+ e^- :\ 7.5\ {\rm fb}, \\
&WW&\to \tau^+ \tau^- \to e^+ e^- :\ 1.2\ {\rm fb} \\
\end{eqnarray*}
We then plot in Fig. 3 dashed contours that mark the boundaries
of the regions where
the selectron signal exceeds background at the $5\sigma$ level. In frame
{\it a}), the reach
can be as high as $m_{\tell_R}=84$ GeV when $m_{\tz_1}$ is as light as
36 GeV, but is diminished when $m_{\tz_1}$ is heavier (large values
of $m_{1/2}$).

{\it 3.2 Charginos:}
Charginos are best searched for in the
mixed hadronic-leptonic final state, {\it e.g.} $\tw_1\overline{\tw_1}\to
\ell\nu_\ell\tz_1 +q\overline{q'}\tz_1$. Grivaz has suggested the set of cuts
(2.4)--(2.9) as optimizing signal to background. Using these cuts, he estimates
a background from $WW$ and other SM processes of 9 fb. We map out the
region yielding a $5\sigma$ signal above background as the dot-dashed contour
in Fig. 3{\it a}-{\it d}. For large values of $m_0\sim 500$ GeV,
LEP2 at $\sqrt{s}=175$ GeV can probe to $m_{\tw_1}=86.7$ GeV in frame {\it b}).
However, for low values of $m_0$, in frames {\it a}), {\it c}) and {\it d}),
the reach
via chargino searches cuts off due to turn on of the two body decay
$\tw_1\to\tnu_{\ell L}\ell$, which then dominates the branching fraction.

{\it 3.3 Neutralinos:}
The background from $WW\to\ell^+\ell^- +\eslt$ forces a more severe set of
cuts to search for $e^+e^-\to\tz_1\tz_2\to\ell\overline{\ell}+\eslt$ events.
To reduce $WW$ background, we require
\begin{eqnarray}
&10&< E_\ell <40\ {\rm GeV},\ \ \ |\eta_{\ell}|<2.5, \\
& \eslt &> 9\ {\rm GeV},\\
& \esl &> 105\ {\rm GeV},\\
& m&(\ell\overline{\ell}) < 50\ {\rm GeV}, \\
& \msl &> 90\ {\rm GeV}, \\
&\phi &(\ell_1 ,\ell_2 ) < 172^o.
\end{eqnarray}
where $\msl$ is the missing mass defined by
$\msl^2 =\esl^2-(\Sigma\vec{p})^2$, where the sum is over observed
particles. The resulting SM background from $WW$ production is
\begin{eqnarray*}
&WW&\to e^+ e^-,\mu^+\mu^- :\ 14.4\ {\rm fb}, \\
&WW&\to \tau^\pm \ell^\mp \to \ell^+ \ell^- :\ 19.2\ {\rm fb}, \\
&WW&\to \tau^+ \tau^- \to \ell^+ \ell^- :\ 3.7\ {\rm fb}, \\
\end{eqnarray*}
where $\ell$ is summed over $e$ and $\mu$. The resulting $5\sigma$ contours
are plotted as dotted lines in Fig. 3. Comparing Fig. 3{\it a} with
Fig. 2{\it a} shows that LEP2 operating at $\sqrt{s}=175$ GeV will
actually
have a somewhat smaller reach for the neutralino dilepton signal
than LEP2 operating at $\sqrt{s}=150$ GeV-- a consequence of the $WW$
background. In addition, for the $\tan\beta =10$ case illustrated in
Fig. 3{\it c}, there is now {\it no} reach for neutralinos in the
dilepton channel.

The search for $\tz_1\tz_2\to$ jets$+\eslt$ is more complicated at
$\sqrt{s}=175$ GeV than at $\sqrt{s}=150$ GeV due not only to the $WW$
background, but also from a background due to $ZH_\ell$ production.
We require
\begin{eqnarray}
&m&({\rm detected})<60\ {\rm GeV},\\
&\eslt &>9\ {\rm GeV},\\
&\esl &>110\ {\rm GeV},\\
&\msl &>90\ {\rm GeV},\\
&\phi &(j_1 ,j_2 ) < 172^o\ ({\rm di-jet\ events}).
\end{eqnarray}
For the $\tan\beta =2$, $\mu <0$ case of Fig. 3{\it a}, we find essentially
no observable region. This is due to combined backgrounds from
{\it i}) $\tw_1\overline{\tw_1}$ production, {\it ii}) $WW$ production, and
mainly {\it iii}) $ZH_\ell\to \nu\overline{\nu}b\overline{b}$ (see Fig. 1{\it
b}).
As for LEP2 at $\sqrt{s}=150$ GeV, there is again no neutralino jets$+\eslt$
signal visible in Fig. 3{\it b}. For the case of Fig. 3{\it c}, the light Higgs
scalar is too massive to be produced, and the main background comes from
\begin{eqnarray*}
WW\to q\overline{q'}\tau\nu_{\tau}\ :\ 6.0\ {\rm fb}.
\end{eqnarray*}
The observable region, delineated by a dashed contour, lies just beyond the
reach for observation of $\tw_1\overline{\tw_1}$. Likewise, a small region
of observability is seen in Fig. 3{\it d}.

{\it 3.4 Higgs bosons:}
An important consequence of the minimal SUGRA model is that the masses
and couplings of the various Higgs bosons are correlated with the masses and
couplings of all the rest of the supersymmetric particles, as well as with the
top quark. In particular, in minimal SUGRA with large $|\mu |$ due to
radiative electroweak symmetry breaking, the lightest Higgs scalar, $H_{\ell}$,
is very much like a SM Higgs boson, but with mass bounded by
$m_{H_\ell}\alt 130$ GeV. Hence, the search for the light Higgs via
$e^+e^-\to ZH_{\ell}$ can explore regions of the same $m_0\ vs.\ m_{1/2}$
plane that can be explored by the search for various SUSY particles.

The search for $e^+e^-\to ZH_{\ell}\to Zb\overline{b}$ proceeds along the same
lines as the search for a SM Higgs, where $H_{SM}\to b\overline{b}$.
Simulations have been carried out for signal and background in a SM Higgs
search (see Ref. \cite{SOPCZAK}), where a discovery cross section at $3\sigma$
for $e^+e^-\to ZH_{SM}$ of $200$ fb was found. We convert this number to a
$5\sigma$ limit for $\int {\cal L}dt=500$ pb$^{-1}$, and take into account
possible variations in the SUSY Higgs production cross section and decay
branching ratio. The resulting dot-dot-dashed contour is
plotted in Fig. 3{\it a}, which probes to $m_{H_{\ell}}\simeq 82$ GeV.
We see that by far the largest region of parameter
space for this case can be scanned via the search for Higgs bosons. If,
however,
a Higgs signal is found, it would be difficult to distinguish in this case
whether it is a SM or SUSY Higgs boson. The $\tan\beta =2$, $\mu >0$ case
illustrated in Fig. 3{\it b} has in general heavier $H_\ell$
than the case
of Fig. 3{\it a}, and so the $H_\ell$ is visible in a much smaller region.
In fact, in Fig. 3{\it b}, the region of Higgs observability occurs when
$ZH_{\ell}\to \ell^+\ell^- +\tz_1\tz_1$ occurs at the 10 event level,
and the Higgs itself is dominated by invisible decay modes to $\tz_1$ pairs.
Finally, in Fig. 3{\it c} and 3{\it d},
none of the $m_0\ vs.\ m_{1/2}$ plane can be
explored via Higgs searches, due to the $H_\ell$ being too massive.

\section{Detecting sparticles at $E_{CM}=190$ and $205$ GeV.}

If LEP2 is operated at $\sqrt{s}=190$ GeV, a smaller total sample of integrated
luminosity $\sim 300$ pb$^{-1}$ is expected to be gathered. In addition, the
SM $WW$ production cross section will increase from 17.3 pb to 19.2 pb, and the
threshold for producing real $ZZ$ events will be passed. The latter are
expected to occur with a cross section of 1.1 pb.

{\it 4.1 Selectrons:}
To evaluate selectron pair production signals at LEP2 at $\sqrt{s}=190$ GeV,
we again use the cuts (3.1-3.4), except for increasing the lepton
energy upper limit to $E_\ell<50$ GeV, and increasing the $\eslt$ cut to
$\eslt >9.5$ (10.3) for $\sqrt{s}=190$ (205) GeV. The background from $ZZ$
production
is again computed using HELAS. The resultant SM backgrounds
for di-electron$+\eslt$ events are
\begin{eqnarray*}
&WW&\to e^+ e^- :\ 10.4\ {\rm fb}, \\
&WW&\to \tau^\pm e^\mp \to e^+ e^- :\ 6.1\ {\rm fb}, \\
&WW&\to \tau^+ \tau^- \to e^+ e^- :\ 0.8\ {\rm fb} \\
&ZZ&\to \nu\overline{\nu}\tau^+ \tau^- \to e^+ e^- :\ 0.2\ {\rm fb}. \\
\end{eqnarray*}
The resultant background at $\sqrt{s}=190$ GeV is smaller than the
corresponding
background for $\sqrt{s}=175$ GeV due to sharper distributions in the
forward region for the higher energy option.
The $5\sigma$ region of observability is plotted in Fig. 4, and
indicated again by the dashed contours. We find that selectron masses of
$m_{\tell_R}\simeq 83-88$ GeV can be probed, depending on the mass and
composition of $\tz_1$.

{\it 4.2 Charginos:}
For chargino pair production at LEP2 at $\sqrt{s}=190$ GeV, we again use
the cuts (2.4-2.9), with the background scaled to the appropriate energy
and luminosity. The regions for chargino discovery via the mixed
hadronic/leptonic event structure are indicated by dot-dashed contours
in Fig. 4. The corresponding reach in terms of $m_{\tw_1}$ increases to
$m_{\tw_1}\sim 94$ GeV for large $m_0$, which is almost at the kinematic limit.

{\it 4.3 Neutralinos:}
To search for $e^+e^-\to\tz_1\tz_2\to\ell\overline{\ell}+\eslt$ events,
we require (after some optimization)
\begin{eqnarray}
&6&< E_\ell <54\ {\rm GeV},\ \ \ |\eta_{\ell}|<2.5, \\
& \eslt &> 9.5\ {\rm GeV},\\
& \esl &> 122\ {\rm GeV},\\
& m&(\ell\overline{\ell}) < 50\ {\rm GeV}, \\
& \msl &> 106\ {\rm GeV}, \\
&\phi &(\ell_1 ,\ell_2 ) < 172^o.
\end{eqnarray}
The SM backgrounds from $WW$ and $ZZ$ production are
\begin{eqnarray*}
&WW&\to \ell^+ \ell^- :\ 20.2\ {\rm fb}, \\
&WW&\to \tau^\pm \ell^\mp \to \ell^+ \ell^- :\ 30.6\ {\rm fb}, \\
&WW&\to \tau^+ \tau^- \to \ell^+ \ell^- :\ 6.0\ {\rm fb}, \\
&ZZ&\to \nu\bar{\nu}\ell^+ \ell^- :\ 0\ {\rm fb},\
({\rm due\ to\ cuts}\ (4.4)\ {\rm and}\ (4.5)), \\
&ZZ&\to \nu\overline{\nu}\tau^+ \tau^- \to \ell^+ \ell^- :\ 0.3\ {\rm fb}, \\
\end{eqnarray*}
where $\ell$ is summed over $e$ and $\mu$. The $5\sigma$ contours
are plotted as usual as dotted lines in Fig. 4.
We see in Fig. 4{\it a} that the dilepton signal from $\tz_1\tz_2$
production yields only a tiny region beyond that which is
explorable via chargino searches, at $\sqrt{s}=190$ GeV. In Fig. 4{\it b},
{\it c} and {\it d}, only a handful of points yielding an observable
dilepton signal were found. These points which fall inside the region
that can be explored via chargino or selectron searches are not shown for
clarity. We thus see that while the neutralino dilepton signal frequently
does not expand the parameter region that might be explored at LEP2, in
favourable cases, it can lead to a confirmatory signal first seen in another
channel.

The search for $\tz_1\tz_2\to$ jets$+\eslt$ is complicated at
$\sqrt{s}=190$ GeV by the fact that there can be a substantial rate for
$ZH_\ell\to\nu\overline{\nu}b\overline{b}$ production over much of
parameter space. For the cases illustrated in Fig. 4{\it a} and 4{\it b},
again, no jets $+\eslt$ signal could be picked out against SM and Higgs
production backgrounds. For the $\tan\beta =10$ case of Fig. 4{\it c},
where the $H_{\ell}$ is still too heavy to be produced,
only a small slice of parameter space yielded a region where the $\tz_1\tz_2$
signal could be seen. To do so, we required
\begin{eqnarray}
&m&({\rm detected})<44\ {\rm GeV},\\
&\eslt &>9.5\ {\rm GeV},\\
&\esl &>126\ {\rm GeV},\\
&\msl &>118\ {\rm GeV},\\
&\phi &(j_1 ,j_2 ) < 172^o\ ({\rm di-jet\ events}).
\end{eqnarray}
Backgrounds from all sources were then negligible, except for
\begin{eqnarray*}
WW\to q\overline{q'}\tau\nu_{\tau}\ :\ 6.1\ {\rm fb}.
\end{eqnarray*}
For Fig. 4{\it d}, no regions of observability for $\tz_1\tz_2\to$ jets$+\eslt$
were found.

{\it 4.4 Higgs bosons:}
For LEP2 at $\sqrt{s}=190$ GeV, we again follow the prescription outlined
in Sec. 3.4 to find regions where $ZH_{\ell}\to Zb\overline{b}$ is detectable,
except for updating the machine energy and luminosity. For Fig. 4{\it a},
the whole of the $m_0\ vs.\ m_{1/2}$ plane shown may be explored via the
Higgs search. The discovery limit contour actually occurs around
$m_{1/2}\sim 300-400$ GeV (shown later in Fig. 6), corresponding to
Higgs masses of $m_{H_\ell}\simeq 93$ GeV. In Fig. 4{\it b}, there now exists
a region of $H_{\ell}\to b\overline{b}$ observability, indicated by the
area {\it between} the dot-dot-dashed contours.
Furthermore, $H_{\ell}\to\tz_1\tz_1$
is detectable below the triple-dot-dashed contours, as in Fig. 3{\it b}.
The small area between these two regions of observability is where the
$H_{\ell}$ branching fraction is split up between the $b\overline{b}$ and
invisible $\tz_1\tz_1$ modes. Here the Higgs signal is just slightly below our
criteria for observability in either mode.
However, if these criteria are relaxed slightly
(to {\it e.g.} a $4\sigma$ effect), or if the luminosity is increased,
then the gap region will become observable. Finally, for the $\tan\beta =10$
case shown in Fig. 4{\it c} and 4{\it d}, the $H_\ell$ is again too heavy
($m_{H_\ell}\agt 95$ GeV) to be seen anywhere in the plane shown.

For completeness, we show in Fig. 5{\it a}-5{\it d} the corresponding region
detectable by LEP2 operating at $\sqrt{s}=205$ GeV and integrated luminosity
of 300 pb$^{-1}$ which has been proposed as a possible upgrade option
for LEP2, particularly for extending the $H_{\ell}$ reach.
Since no new SM backgrounds open up, we use the same cuts
as in the $\sqrt{s}=190$ GeV case. In general, the various search regions
expand somewhat from the $\sqrt{s}=190$ GeV plot of Fig. 4. The main
difference comes in the search for the light Higgs boson. In Fig. 5{\it a},
the LEP2 reach for Higgs bosons at $\sqrt{s}=205$ GeV has expanded to a
contour at around $m_{1/2}\sim 500-600$ GeV. Fine-tuning arguments\cite{FINE}
would suggest that such a reach (which corresponds to a
gluino (chargino) mass of $\sim 1400$ ($\sim 500$)~GeV) essentially probes
all of the parameter space of weak scale supersymmetry. Such a conclusion
should be viewed in perspective. First, the fine-tuning criteria are
subjective. Second, as shown below, the range in
$m_{1/2}$ explorable via the Higgs boson search; {\it i.e}
the correlation between
$m_{H_{\ell}}$ and the gaugino mass is very sensitive to other parameters.
For instance, for the $\tan\beta =2$
$\mu >0$ case of Fig. 5{\it b}, the Higgs reach has extended to around
$m_{1/2}\sim 250$ GeV, although the slight gap of difficult observability
persists around $m_{1/2}\sim 100-120$ GeV.
Furthermore, LEP2 at $\sqrt{s}=205$ GeV
finally has a significant reach for the high $\tan\beta$ case of Fig. 5{\it c},
where $e^+e^-\to ZH_{\ell}$ can now be seen to $m_{1/2}\sim 100$ GeV, for
$m_0<200$ GeV. The $\tan\beta =10$, $\mu >0$ case of Fig. 5{\it d} still has no
region of Higgs observability, since $m_{H_{\ell}}>100$ GeV throughout the
allowed plane. We thus conclude that while the increased energy of LEP2
substantially expands the parameter space region that can be explored via
the Higgs boson search, non-observation of any signal cannot unequivocally
exclude even this very restricted
framework even if LEP2 is operated at 205~GeV. Of course, the observation
of the Higgs signal alone, while very welcome, would not serve to distinguish
the SUSY framework from the SM.

\section{Other multi-lepton signals}

In addition to the more typical signals for supersymmetry at LEP2
already discussed, there exist additional signals which have rarely been
addressed in the literature. These include signals containing three or four
isolated leptons. For example, $3\ell +$ jets$+\esl$ can come from
\begin{itemize}
\item $\tnu_{\ell}\overline{\tnu_{\ell}}$ production, where
$\tnu_{\ell}\to \nu_{\ell}\tz_2\to \nu_{\ell}e\overline{e}\tz_1$, whereas
$\overline{\tnu_{\ell}}\to e\tw_1\to eq\overline{q'}\tz_1$.
\end{itemize}
Likewise, $4\ell +\esl$ events can come from
\begin{itemize}
\item $\te_i\overline{\te_j}$ ($i,j=L$ or $R$) production, where, for instance,
$\te_i\to\tz_2 e$ followed by $\tz_2\to e\te_R\to e\overline{e}\tz_1$, while
the original $\te_j\to e\tz_1$,
\item $\tz_2\tz_2\to \ell\overline{\ell}\tz_1 +\ell'\overline{\ell'}\tz_1$, and
\item $\tnu_{\ell}\overline{\tnu_{\ell}}$ production, where sneutrinos
decay via $\tnu_{\ell}\to \nu_{\ell}\tz_2\to \nu_{\ell}e\overline{e}\tz_1$
or $\tnu_{\ell}\to \ell\tw_1\to \ell \ell'\nu_{\ell'}\tz_1$.
\end{itemize}
The above reactions generally occur within subsets of the regions of
parameter space already delineated in Sec. II-IV, and so give no
additional reach for supersymmetry.
The detection of such events is nonetheless important since it could serve
to test the details of the underlying model.

As an example, in Fig.~6
we delineate regions of parameter space where $4\ell +\esl$
events are observable, assuming the $\sqrt{s}=190$ GeV option for LEP2.
We require
\begin{eqnarray}
& E_\ell & > 3\ {\rm GeV},\ \ \ |\eta_{\ell}|<2.5, \\
&\eslt &> 9.5\ {\rm GeV},
\end{eqnarray}
and then require at least 5 signal events for observability, since SM
backgrounds should be tiny because $ZZ $ production and direct
decays to $e$ or $\mu$ pairs can be easily vetoed,
and $ZZ$ or $ZH_\ell \rightarrow 4\tau \rightarrow 4\ell$ cross section is
small.
The resulting regions are plotted in
Fig. 6{\it a}-6{\it d}. In Fig. 6{\it a}, there is unfortunately only
a tiny region of $4\ell$ observability, while in Fig. 6{\it b}
there is a substantial region, outlined by the dashed contour. Within the
dashed contour, the dotted contours delineate which reaction
dominantly contributes to the signal. In Fig. 6{\it b}, $\te_L\te_R$
and $\tz_2\tz_2$ are the dominant production mechanisms over most of the
observable region, with $\tnu_{\ell}\overline{\tnu_{\ell}}$ contributing
dominantly in a smaller region. For the $\tan\beta =10$ cases of Fig. 6{\it c}
and 6{\it d}, $\tnu_{\ell}\overline{\tnu_{\ell}}\to 4\ell$ dominates for
$60< m_0 <100$ GeV. For reasons of brevity, we do not show regions where
trilepton signals occur at observable rates, nor do we show the energy
dependence of the $3\ell$ and $4\ell$ signals.

\section{Comparison of various LEP2 energy upgrade options and
comparison with Tevatron MI}

We show in Fig. 7 the cumulative search contours for LEP2 energies of
$\sqrt{s}=175$, 190 and 205 GeV, with respective integrated luminosities
of $\int {\cal L}dt=500$, 300 and 300 pb$^{-1}$. The contours are
composites of those shown in Figs. 3-5, with some tiny additional regions
added in where, for instance, overlapping slepton and neutralino signals
can increase the SUSY discovery reach. It is clear to see that
the energy increase from $\sqrt{s}=175$ GeV, to 190 and 205 GeV results
in increased detectability for charginos from roughly 87 GeV, to 95 GeV
and 102 GeV, respectively (for $m_0$ large). Likewise, in the small
$m_0$ region ($m_0\simeq {m_{1/2}\over \sqrt{3}}$), selectron masses
of $m_{\tell_R}\simeq 82$, 88 and 96 GeV can be probed. This is a clear
argument for LEP2 to try to attain the highest energy option. An exception
to this does occur, however, for observation of the $\tz_1\tz_2$
reaction. As can be seen in the $m_0\sim 100$ GeV region of Fig. 7{\it a},
all three (and even the $\sqrt{s}=150$ GeV option)
energy upgrades have roughly equivalent reach. This is due to the
fact that LEP2 operating at a reduced energy can have a similar or perhaps
even better chance for observing neutralino pairs than the higher energy
options. This situation occurs because $\tz_1\tz_2$ production has a very
small cross section, and the additional backgrounds from $WW$, $ZZ$ and
$ZH_\ell$ production at higher energies can swamp the tiny neutralino
pair signal. In addition, supersymmetric processes such as
$\tw_1\overline{\tw_1}$ production can mask some of the region
where $\tz_1\tz_2$ might have otherwise been visible.

In Fig. 7, in addition to the cumulative contours for the three LEP2
energy and luminosity options, we have as well plotted the approximate reach of
experiments operating at the Fermilab $p\bar p$ collider in the Main
Injector (MI) era. The Tevatron MI is expected to turn on around 1999,
at $\sqrt{s}=2$ TeV, and it is expected to accumulate $\sim 1$ fb$^{-1}$
of integrated luminosity per year. Recently, the reach in $m_{\tg}$ has been
calculated for searches in the multi-jet$+\eslt$ channel which results
from $p\bar p\to\tg\tg$, $\tg\tq$ and $\tq\tq$ production\cite{LOPEZ,BKT}.
It was found that the Tevatron MI could probe to $m_{\tg}\sim 200-270$ GeV
($m_{\tq}>>m_{\tg}$), or $m_{\tg}\sim 265-350$ GeV ($m_{\tq}\sim m_{\tg}$).
We combine the more optimistic of these values\cite{LOPEZ} with the recent
calculation of the Tevatron MI reach for SUSY via trileptons and dileptons
from chargino/neutralino production\cite{TEVST}
(see also Ref. \cite{LOPEZ,MRENNA}).
The resultant small dashed contours are shown in Fig. 7, and labelled by MI.
For all four cases shown, it is clear that
LEP2 will have a larger reach for minimal SUGRA in the large $m_0$ region,
due to searches for chargino pairs. In addition, LEP2 can probe regions of
small $m_0$ not accessible to the MI, via the search for selectrons.
However, in the intermediate region of $m_0\sim 100-200$ GeV, Tevatron MI
experiments can have a superior reach to LEP2, mainly via the search
for $\tw_1\tz_2\to 3\ell +\eslt$ events.
These contours clearly illustrate
the complementary  capabilities of LEP2 $e^+e^-$ and Tevatron MI $p\bar p$
colliders.

In Fig. 8{\it a} and {\it b}, we show the regions of
$m_0\ vs.\ m_{1/2}$ space explorable via
Higgs searches, for the two $\tan\beta =2$ cases.
For $\tan\beta =10$, $\mu <0$
case, the light Higgs is too heavy to be observed at any of the LEP2
energy/luminosity options except for $\sqrt{s}=205$ GeV, which is plotted
in Fig. 5{\it c}.
For the $\tan\beta =10$, $\mu >0$ case, the light Higgs is too heavy to be
observed at {\it any} of the considered LEP2 energy/luminosity options.
For the $\sqrt{s}=150$ GeV option,
of course, no Higgs signal is visible beyond LEP1 bounds.
In Fig. 8{\it a}, the $\sqrt{s}=175$ GeV
energy option can explore up to $m_{H_{\ell}}\sim 82$ GeV, which covers
a significant portion of the $\tan\beta =2$, $\mu <0$, $A_0=0$
parameter space -- well beyond
the regions for any SUSY particle searches. The modest energy increase to
$\sqrt{s}=190$ GeV considerably increases the reach in the SUGRA space
explorable via the
Higgs search. This is primarily because $H_{\ell}$ cannot become too heavy
within this (and many other) model framework(s).
The $\sqrt{s}=205$ GeV option can probe essentially the whole range of
parameters allowed by fine-tuning arguments\cite{FINE} for negative values
of $\mu$. If $\mu > 0$, the region probed via the Higgs search is
significantly smaller.
We see from Fig. 8{\it b} that the $\sqrt{s}=175$ GeV LEP2 option could
only explore a small region of SUGRA space, and that via the invisible
Higgs signal. The energy increase to $\sqrt{s}=190$ GeV would not increase
the invisible Higgs region significantly, but would probe the interior
of the dot-dashed region via a $ZH_{\ell}\to Zb\overline{b}$ search.
An increase of energy to $\sqrt{s}=205$ GeV would substantially increase
the region seeable via $ZH_{\ell}\to Zb\overline{b}$, but would leave the lower
gap region around $m_{1/2}\sim 100$ GeV still on the edge of observability
in both visible and invisible Higgs boson decay channels.

We conclude that
non-observation of a Higgs signal would rule out a huge region of this
particular plane particularly if $\mu < 0$. It
should, however,
be kept in mind that the detection of the Higgs boson signal would not be
conclusive evidence for SUSY, since $H_{\ell}$ is essentially indistinguishable
from
the SM Higgs boson. Higgs boson detection
via its invisible mode, which is possible for positive values of $\mu$,
would of course imply the existence of a non-SM Higgs sector.

\section{Neutralino search in the SUGRA-inspired MSSM ($|\mu |$ free case)}

All of the preceeding analysis has been performed within the framework of
the minimal SUGRA model which, because of the assumed symmetries
of dynamics at the GUT scale, is determined by just a few
parameters renormalized at around the same scale. The diversity of sparticle
masses and couplings, renormalized at the weak scale relevant for
phenomenology, then arises via renormalization effects when common GUT scale
mass and coupling parameters are evolved down to the weak scale. As discussed
in Sec. I, these same radiative corrections can lead to the
observed pattern of electroweak symmetry breaking, provided GUT scale
parameters are chosen within certain ranges: then, the superpotential Higgs
mass $\mu$ is determined up to a sign, since $\mu^2$ is
tuned to give the correct value of $M_Z$.

Despite the fact that SUGRA models provide an attractive and economic
framework,
it should be kept in mind that any of the underlying assumptions could prove
to be incorrect.
Indeed many theoretical
and experimental analyses have been cast within the framework of the
SUGRA-inspired MSSM framework, where it is generally assumed that the
soft-breaking gaugino masses $M_1$, $M_2$ and $M_3$ are related as in a GUT
model, and that the squark and slepton masses originate from a universal
GUT scale soft-breaking term $m_0$\cite{FN2}. These requirements are easily
implemented via simple formulae relating squark, slepton and gluino
masses. The additional relationships between
the Higgs boson masses, the $A$ parameters, and the electroweak symmetry
breaking requirement (which usually requires a complete,
coupled RGE solution) are then neglected, so that $m_{H_p}$, $A_t$ and
$\mu$ are left as free parameters. The resulting model maintains some of
the important mass relationships contained in minimal SUGRA, but then has
additional parameters, giving it more generality, but also making
parameter space scans more tedious.

What are the main effects of relaxing these conditions for LEP phenomenology?
In
minimal SUGRA, generally $|\mu |>>M_1$, $M_2$ and $M_W$,
which results in gaugino-like
$\tw_1, \tz_1$ and $\tz_2$. Then, the couplings of the two lighter neutralinos
(which are likely to be kinematically accessible at colliders) to the $Z$
boson is strongly suppressed, so that their production by $e^+e^-$ collisions
is also strongly suppressed, particularly when selectrons are heavy (see
Fig.~1).
This is not true for lighter charginos which in fact have enhanced
$SU(2)$ triplet couplings to the $Z$, and which, of course, also couple to
photons. If we allow that we do not know the high scale dynamics, and so,
relax the constraints from radiative electroweak symmetry breaking, $\mu^2$
can be rather small, so that the lighter charginos and neutralinos can be
Higgsino-like. In this case, the neutralino cross section at LEP2
may be larger by orders of magnitude (recall Higgsinos have gauge
couplings to $Z$) relative to the SUGRA case, while that of charginos
and other charged sparticles or sneutrinos is comparatively unaffected.
As a result, neutralino phenomenology may be very different, while the
phenomenology of charginos and sleptons is roughly as discussed
above\cite{FN3}.

Indeed earlier studies within the SUGRA-inspired MSSM framework have
shown that the search for chargino pairs
(and also sleptons) would generally
proceed as discussed earlier for the minimal SUGRA model, and in general,
charginos ought to be visible if their production is kinematically
allowed\cite{OLD1,OLD,FENG}.
The case for neutralino pair production, however,
can be quite different.
When $|\mu |$ is small,
as allowed in the SUGRA-inspired MSSM, then $\tz_1\tz_2$ production can
take place via the $s$-channel $Z$ exchange graph, and the total $\tz_1\tz_2$
production cross section can be comparable to the total $\tw_1\overline{\tw_1}$
cross section, even if selectrons are quite heavy.
These neutralino total cross sections have recently been displayed
in the $\mu\ vs.\ M_2$ plane\cite{MAJEROTTO,MELE}, although without
explicit simulation and background evaluation. Because neutralino production
rates are extremely sensitive to the assumption of the radiative symmetry
breaking mechanism, and because relatively little work has been done on the
prospects of detecting these signals at $e^+e^-$ colliders, we felt it
worthwhile to single out neutralino signals for a closer investigation.
In the interest of brevity, and because this analysis is outside the main
theme of this paper, we illustrate this for just one centre of mass energy,
$\sqrt{s}= 190$~GeV.

To this end, we plot in Fig. 9 the cross section after cuts (see Sec. IV) from
all SUSY and Higgs boson
sources of $\ell\overline{\ell}+\eslt$, $\ell + {\rm jets}+\eslt$ and, finally,
${\rm jets}+\eslt$ events
for the $\sqrt{s}=190$ GeV option of LEP2.
Chargino and neutralino production and subsequent decays are the primary
source of these events in the figure, where
for definiteness, we have chosen
$m_{\tg}=600$~GeV, $m_{\tq}=1000$~GeV (this roughly corresponds to $m_0 \sim
850$~GeV, $m_{1/2} \sim 220$~GeV in the SUGRA case --- a look at Fig.~4
shows that there would be no observable SUSY signal within this framework for
parameters in this range),
and $\tan\beta=2$. The sleptons, whose
masses are fixed by SUGRA relations, have masses $\sim 850$~GeV, and so are too
heavy to be of direct interest. The weak scale $A$ parameters and $m_{H_p}$
are chosen to be -1000~GeV and 500~GeV, respectively, but our results are
insensitive to this choice. The band between the vertical lines is excluded
by the non-observation of any SUSY signal or deviations in the $Z$ line shape
in experiments at LEP. We see that these signals are at or above the
$5\sigma$ level for a substantial region of small $\mu^2$.
Several comments are worthy of note.

\begin{itemize}
\item When $|\mu |$ is large, as is typical in SUGRA models,
$\tw_1\overline{\tw_1}$ and $\tz_1\tz_2$ production is kinematically not
allowed,
but as $|\mu |$ decreases, $m_{\tw_1}$ and $m_{\tz_{1,2}}$ also decrease
until their production becomes kinematically accessible. Single lepton
production can only come from chargino decays (except when a lepton from
the decay of a neutralino escapes undetected) while chargino and neutralino
production can both contribute to the dilepton and dijet signals.

\item For values of $\mu$ in the central region, where the cross sections
are substantial, we see from the relative size of the $1\ell$ and the other
cross sections that chargino and neutralino production indeed
contribute comparable amounts as anticipated.

\item For positive values of $\mu$, the chargino is typically lighter than when
$\mu$ is negative, and has the same magnitude. This is reflected in the
relatively larger single lepton cross sections in this case compared to $\mu
< 0$.

\item The relative
contributions of the various SUSY processes to the dilepton and
dijet signals is sensitive to the parameters. For the choice in the figure,
particularly at the large $|\mu|$ edge where the signals drop rapidly, the
bulk of the neutralino contribution comes from $\tz_1\tz_3$ production: This
is because the $\tz_3$ tends to have a substantially larger Higgsino component
than $\tz_2$ for the particular parameter choice. The cross sections
drop rapidly once $\tz_1\tz_3$ production becomes kinematically forbidden.
At this point $\tz_1\tz_2$ production is still allowed which is why the
cross sections do not go to zero; the smallness of this cross section
reflects the large suppression of the $Z\tz_1\tz_2$ coupling. The cross over
between the single lepton and the dilepton curves for large negative values
of $\mu$ occurs because $\tw_1\overline{\tw_1}$ production becomes
kinematically
inaccessible even when $\tz_1\tz_3$ production continues to remain allowed.

\end{itemize}

We have just seen that if lighter neutralinos are Higgsino-like, they will
lead to characteristic signals at rates comparable to chargino production.
Since these signals can also come from chargino production, it is reasonable
to ask if one can distinguish chargino production alone from chargino and
neutralino production. Several possiblities come to mind. ({\it i})~Neutralino
production does not lead to single lepton topologies except when a lepton
escapes experimental detection: observation of a substantial rate for this
would point to charginos as the source of such events. ({\it ii})~Chargino
pair production is as likely to yield like flavour as unlike flavour dilepton
pairs. An excess of $e^+e^-$ and $\mu^+\mu^-$ relative to $e^{\pm}\mu^{\mp}$
events would likely indicate the simultaneous
production of charginos along with sleptons or neutralinos. Slepton
production generally occurs at a significantly larger rate and (unless
sleptons can also decay to charginos) does not result in jet(s) plus $\eslt$
events. ({\it iii})~The mass of $\ell^+\ell^-$ pairs from $\tz_i\tz_1$
production
is bounded above by $m_{\tz_i} -m_{\tz_1}$ while the corresponding
distribution from chargino pair production is expected to be much broader.
Unfortunately, this does not always prove to be useful. We have checked, for
example, that for positive values of $\mu$ where the chargino is
significantly lighter than $\tz_2$ the distributions look crudely similar,
and so may be difficult to distinguish. For $\mu < 0$ this distinction
might well be possible.

To sum up the results of this section, neutralino production rates are small
in supergravity models with radiative symmetry breaking unless sleptons are
also light. If, however, we relax the radiative symmetry breaking ansatz,
neutralino production may be increased by orders of magnitude, and an
observable signal might be possible. In this case, it would be interesting
to see if neutralino events can be distinguished from chargino events which
would occur at similar rates.

\section{Summary}

We have used ISAJET to examine
signals for various supersymmetric processes that might be
accessible at LEP2. Our study differs from most earlier LEP2 studies in that we
work within the economic minimal SUGRA framework which is fixed in terms of
just four parameters along with the sign of the superpotential Higgsino mass
parameter ($\mu$). This introduces correlations between various
sparticle masses which, in turn, lead to new features
(that are absent in the
supergravity-inspired MSSM framework)
in the behaviour of different
cross sections as a function of model parameters.
Also, several SUSY reactions may be kinematically accessible at the same time.
We have devised cuts to separate
SUSY and Higgs boson signals from SM backgrounds, and also to differentiate
as much as possible between various signal processes.

Our purpose has been
to assess the reach of LEP2 and compare four different options for an energy
upgrade. The details of our computations are given in Secs. II-IV, and
in Figs. 2-5 where the regions of the $m_0-m_{1/2}$ plane that might be probed
via different channels are shown. Generally speaking, increasing the
center-of-mass energy
increases the reach in $m_{1/2}$ by about (50-70)\% of the increase in energy
as illustrated in Fig.~7, where the composite region of the plane that can
be probed via any SUSY channel is shown.
There is, however, an important exception: if LEP2 is run below the $WW$
threshold, SM backgrounds are greatly reduced so that the $\tz_1\tz_2$
reaction might be more easily seen than at LEP2 operating at higher
energies.
Fig.~7 also shows for comparison the reach of the
Main Injector upgrade of the Tevatron. We see that there are regions of
parameter space where the Tevatron upgrade significantly outperforms even
the highest energy option considered for LEP2, while in other regions, exactly
the opposite is true. This illustrates the complementarity between $e^+e^-$
and hadron colliders.

Unlike the case of sparticle searches, where increasing LEP2 energy leads
to modest increase in the SUSY reach of the machine,
a relatively small increase in the machine
energy results in a significantly larger reach when probing the Higgs sector.
This stems from the well known fact that the mass of the lightest neutral
Higgs scalar is bounded by $\sim 130$~GeV within this framework, and is
illustrated in Fig.~8. A definitive
non-observation of {\it any} Higgs boson signal
(including the missing energy signal from $H_{\ell} \rightarrow \tz_1\tz_1$
decays)
would result in stringent restrictions on the model parameters. Unfortunately,
however, while
the observation of a Higgs boson signal in one of the SM modes would be
most welcome, it would not (unless we are extremely lucky) serve to
distinguish the SUSY framework from the SM. From this point of view, there
appears to be no substitute for a direct observation of sparticles.
Motivated by the fact that neutralino production rates can be very
different from their SUGRA model values if they contain substantial Higgsino
components, and that the mechanism for electroweak symmetry
breaking is essentially unknown, in Sec. VII
we relax the radiative symmetry breaking constraint
which fixes $|\mu|$ to be large, and forces the lighter neutralinos
to be gaugino-like. Our main results are illustrated in Fig.~9,
where it is shown that neutralino cross sections could become comparable
to those of charginos, although there are only small regions of
parameter space where neutralino signals might be observable and where
chargino pair production is kinematically forbidden.

In summary, we have examined the prospects for the detection of supersymmetry
at various energy upgrade options of LEP2 within the framework of the
minimal supergravity model. Because of the underlying correlations between
sparticle masses, different reactions probe different regions of the
parameter space. The Fermilab Main Injector upgrade that is expected to
become operational just before the turn of the century is complementary
to LEP2 upgrades in that both facilities can probe significant ranges of
parameters where there are no observable signals at the other facility.


\acknowledgments

We thank M. Corden and C. Georgiopoulos for discussions, and
G. Giudice and M. Mangano of the LEP2 SUSY working group for
discussions and motivation.
HB thanks U. Nauenberg and the University of Colorado at Boulder
for hospitality and motivation while upgrading ISAJET.
This research was supported in part by the U.~S. Department of Energy
under grant numbers DE-FG-05-87ER40319 and DE-FG-03-94ER40833.

%

%
\newpage

\begin{figure}
\caption[]{Total cross sections versus common GUT scale gaugino mass
$m_{1/2}$ for various particle creation mechanisms
within the minimal SUGRA model, for $e^+e^-$ reactions at $\sqrt{s}=175$ GeV.
We take $A_0=0$, $\tan\beta =2$ and $\mu <0$.
In {\it a}), we show results for $m_0=0$, while in {\it b}) we show
results for $m_0=m_{1/2}$, and in {\it c}) we take $m_0=4m_{1/2}$.
}
\end{figure}

\begin{figure}
\caption[]{Regions of the $m_0\ vs.\ m_{1/2}$ plane explorable at LEP2 with
$\sqrt{s}=150$ GeV, and $\int {\cal L}dt=500$ pb$^{-1}$. In all frames, we
take $A_0=0$. In {\it a}), we take $\tan\beta =2$, $\mu <0$, while
in {\it b}) we take $\tan\beta =2$ with $\mu >0$.
In {\it c}), we take $\tan\beta =10$, $\mu <0$ and in {\it d}) we take
$\tan\beta =10$, $\mu >0$.
The regions denoted
by TH are excluded by theoretical constraints, while the region labelled EX
is excluded by experimental constraints.
}
\end{figure}
\begin{figure}
\caption[]{Same as Fig. 2, except for $\sqrt{s}=175$ GeV, and
$\int {\cal L}dt=500$ pb$^{-1}$.
}
\end{figure}
\begin{figure}
\caption[]{Same as Fig. 2, except for $\sqrt{s}=190$ GeV, and
$\int {\cal L}dt=300$ pb$^{-1}$.
}
\end{figure}
\begin{figure}
\caption[]{Same as Fig. 2, except for $\sqrt{s}=205$ GeV, and
$\int {\cal L}dt=300$ pb$^{-1}$.
}
\end{figure}
\begin{figure}
\caption[]{A plot similar to Fig. 4 (for $\sqrt{s}=190$ GeV, and
$\int {\cal L}dt=300$ pb$^{-1}$), except we show regions yielding at least
5 events containing four isolated leptons that do not come from
$Z$ pairs. The complete region where there are $4\ell$ signals from
all sources is outlined
by the dashed contours; the dotted contours delineate regions where most
or all of the signal comes from the particular reaction shown on the figure.
}
\end{figure}
\begin{figure}
\caption[]{Cumulative reach of various LEP2 upgrade options for
supersymmetric particles (excluding Higgs bosons), for
$\sqrt{s}=175$ GeV and $\int {\cal L}dt=500$ pb$^{-1}$ (dashed),
$\sqrt{s}=190$ GeV and $\int {\cal L}dt=300$ pb$^{-1}$ (dot-dashed), and
$\sqrt{s}=205$ GeV and $\int {\cal L}dt=300$ pb$^{-1}$ (dotted). Also shown
for comparison is the combined reach of Tevatron Main Injector era
experiments ($\sqrt{s}=2$ TeV and  $\int {\cal L}dt=1000$ pb$^{-1}$)
(dashed curve labelled by MI).
}
\end{figure}
\begin{figure}
\caption[]{Reach of various LEP2 upgrade options for the lightest Higgs
boson $H_\ell$, for the $\sqrt{s}=175$ GeV (dashes), $\sqrt{s}=190$ GeV
(dot-dashed) and
$\sqrt{s}=205$ GeV cases (dots), with an integrated luminosity
as in Fig.~7 for $\tan\beta =2$ and {\it a}) $\mu <0$
and {\it b}) $\mu >0$.
}
\end{figure}
\begin{figure}
\caption[]{Cross section for various event topologies for LEP2 at
$\sqrt{s}=190$ GeV, after the cuts of
Sec. IV, for supersymmetric signals versus the $\mu$ parameter. Here, $\mu$
is taken as a free parameter because
the requirement of radiative electroweak symmetry breaking has been dropped.
}
\end{figure}


\begin{references}
\bibitem{FN1} For a gaugino-like chargino, the ALEPH experiment
has obtained a lower bound of 47~GeV, slightly better than the general bound
cited in Sec. 1.
\bibitem{FN2} In principle, each of the soft gaugino masses as well as
the soft SUSY breaking masses of each $SU(3) \times SU(2) \times U(1)$
multiplet should be taken as an independent parameter. The resulting
proliferation of parameters would make phenomenological analyses impossible.
The SUGRA GUT inspired assumptions for these masses make these analyses
tractable and serve to test the sensitivity of the resulting signals to at
least some of the GUT scale assumptions.
\bibitem{FN3} Cross sections of sparticles with non-trivial electroweak
quantum numbers are changed by factors of order unity by changes of mixing
angle. Also, there can be kinematic effects because of differences in
sparticle masses. An important effect for chargino phenomenology in the
Higgsino region is that for $M_1, M_2 >> \mu, M_W$, $m_{\tw_1} \simeq
m_{\tz_1} \simeq m_{\tz_2}$, so that the decay products of charginos and
neutralinos tend to be rather soft.
\end{references}
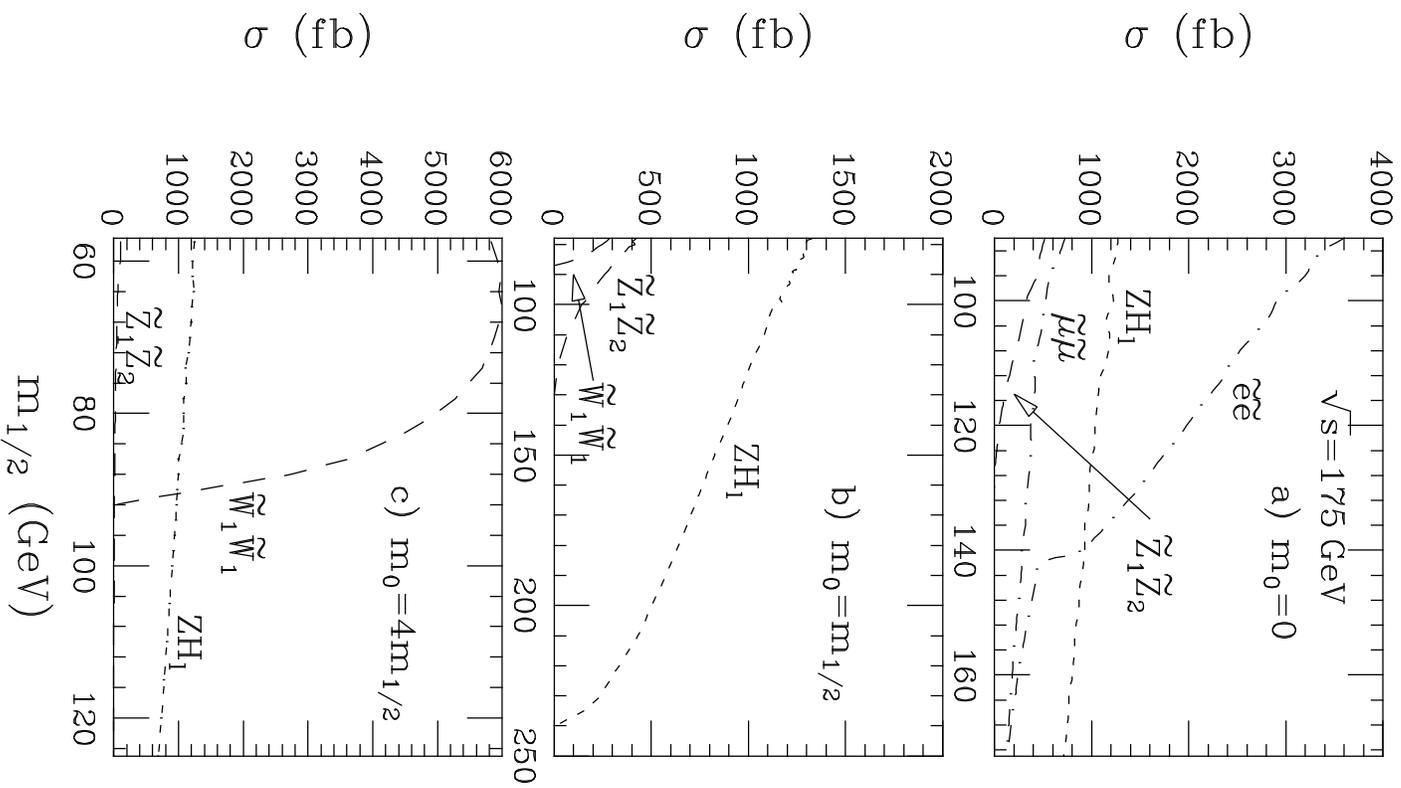
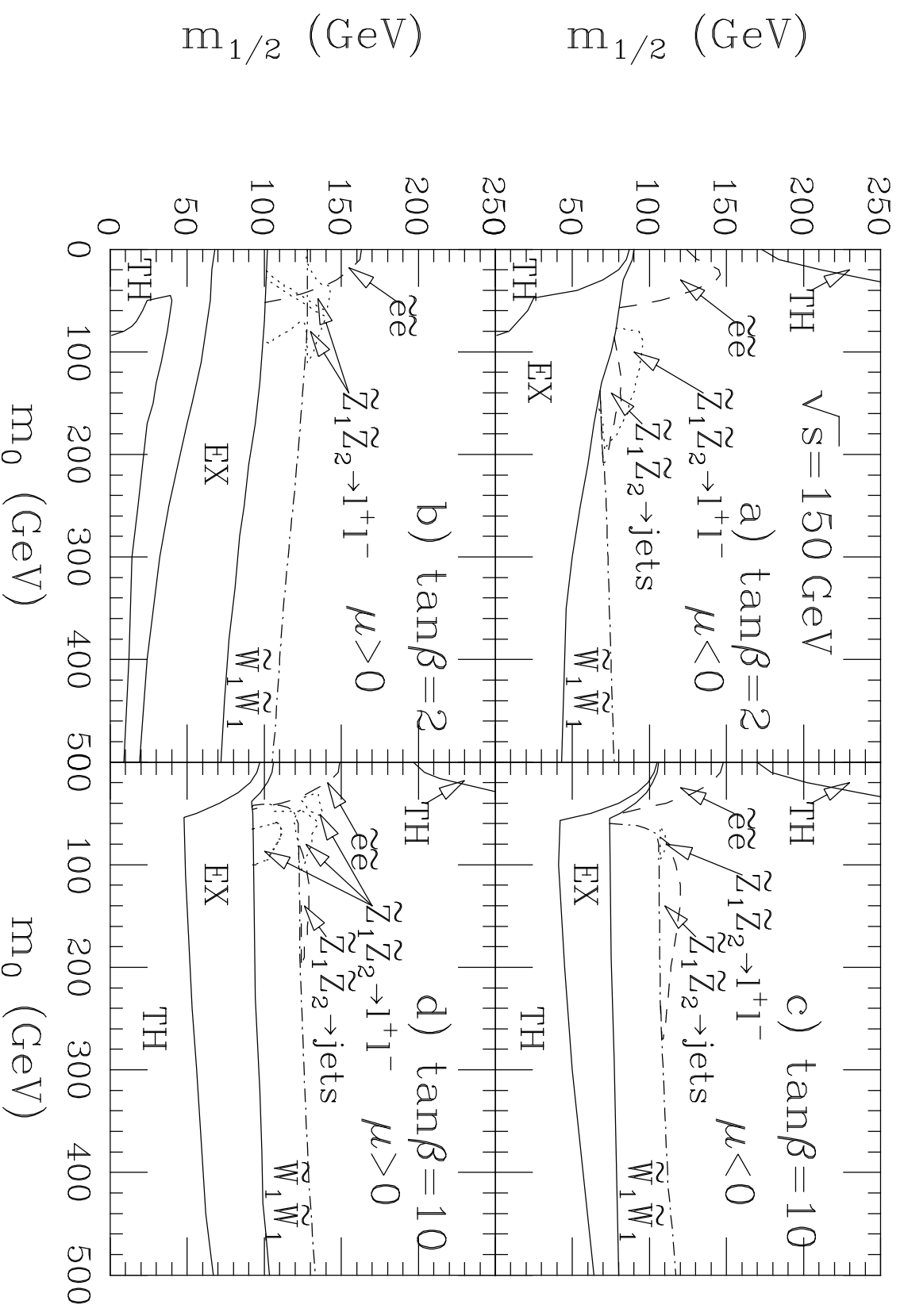
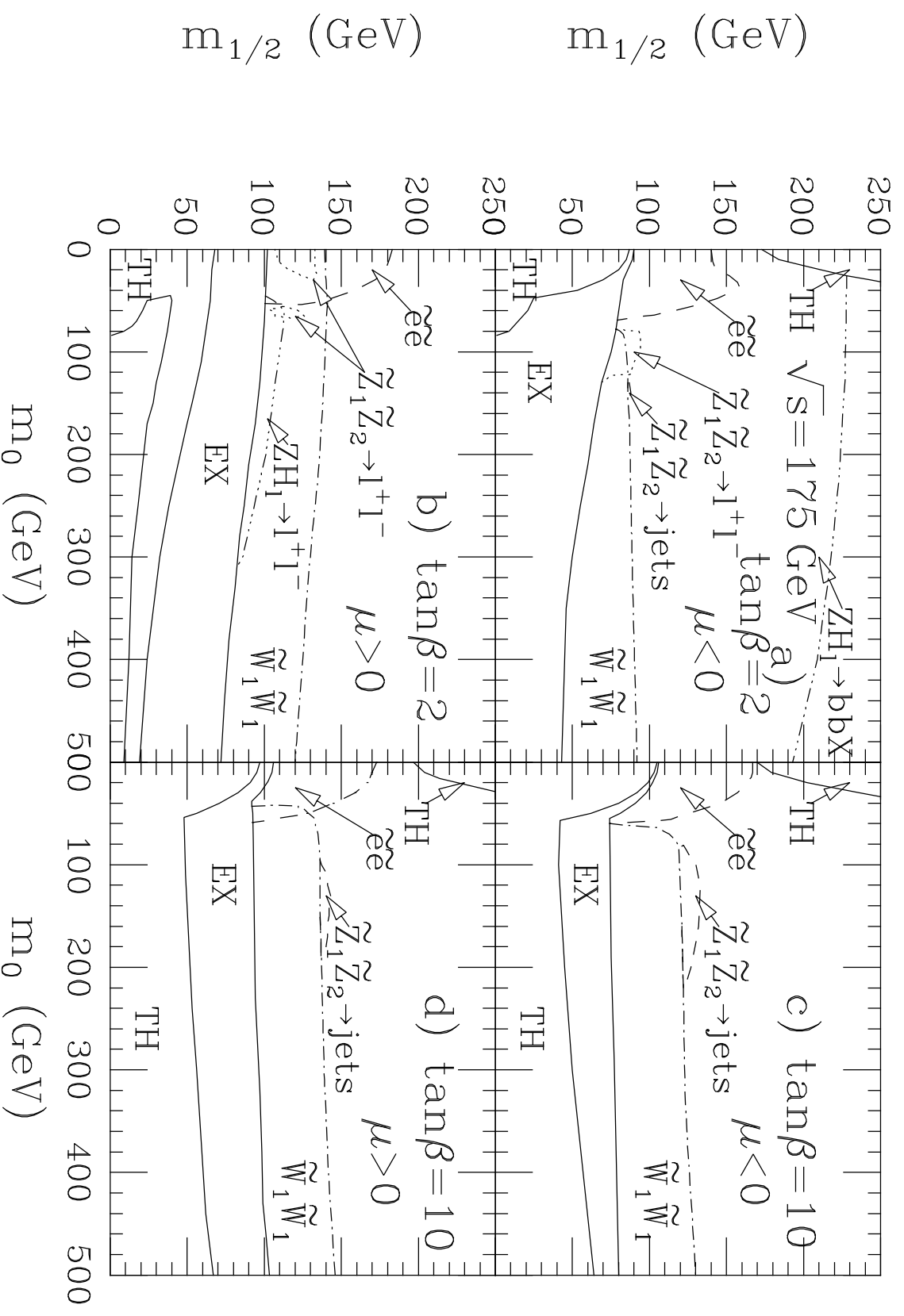
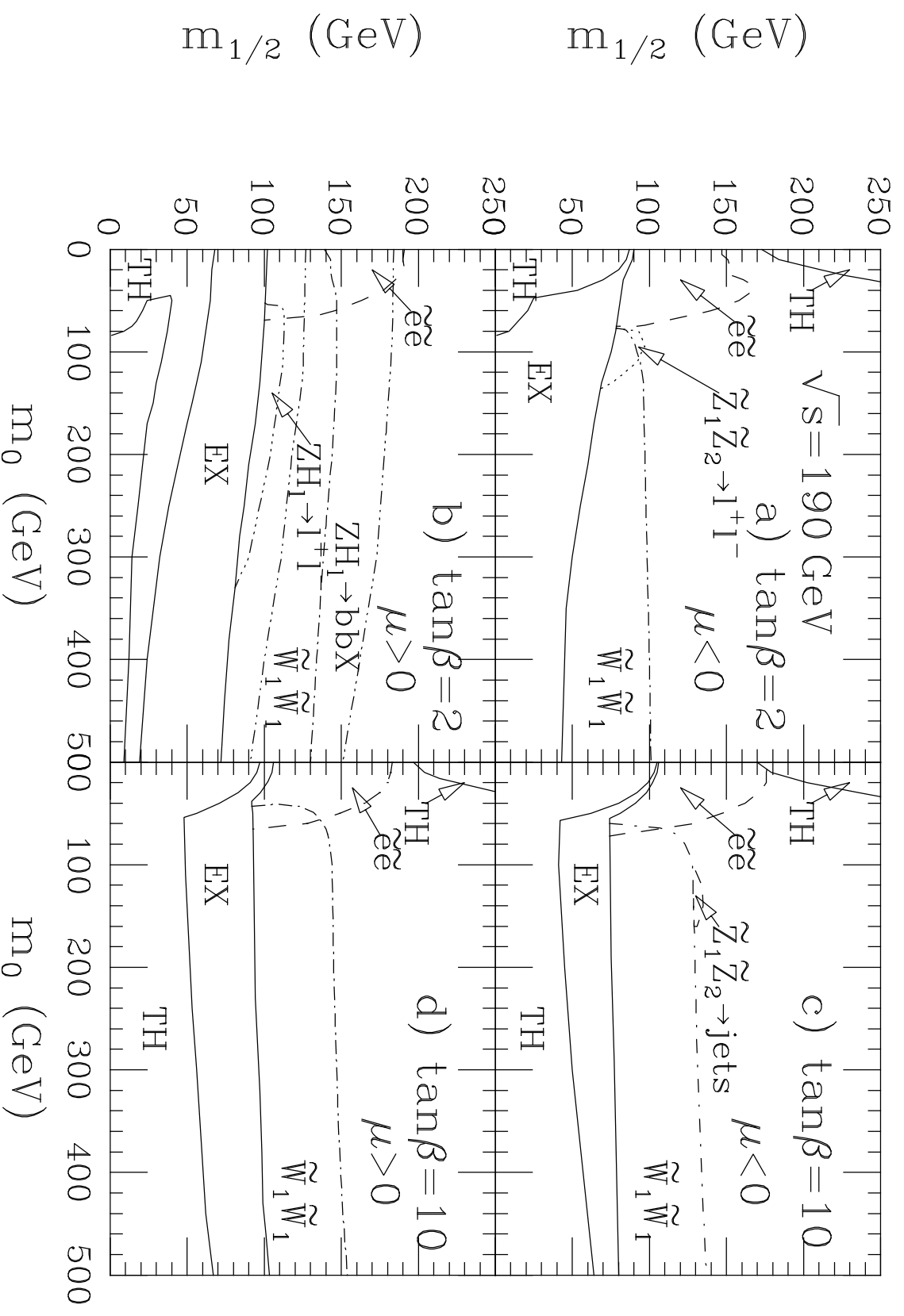
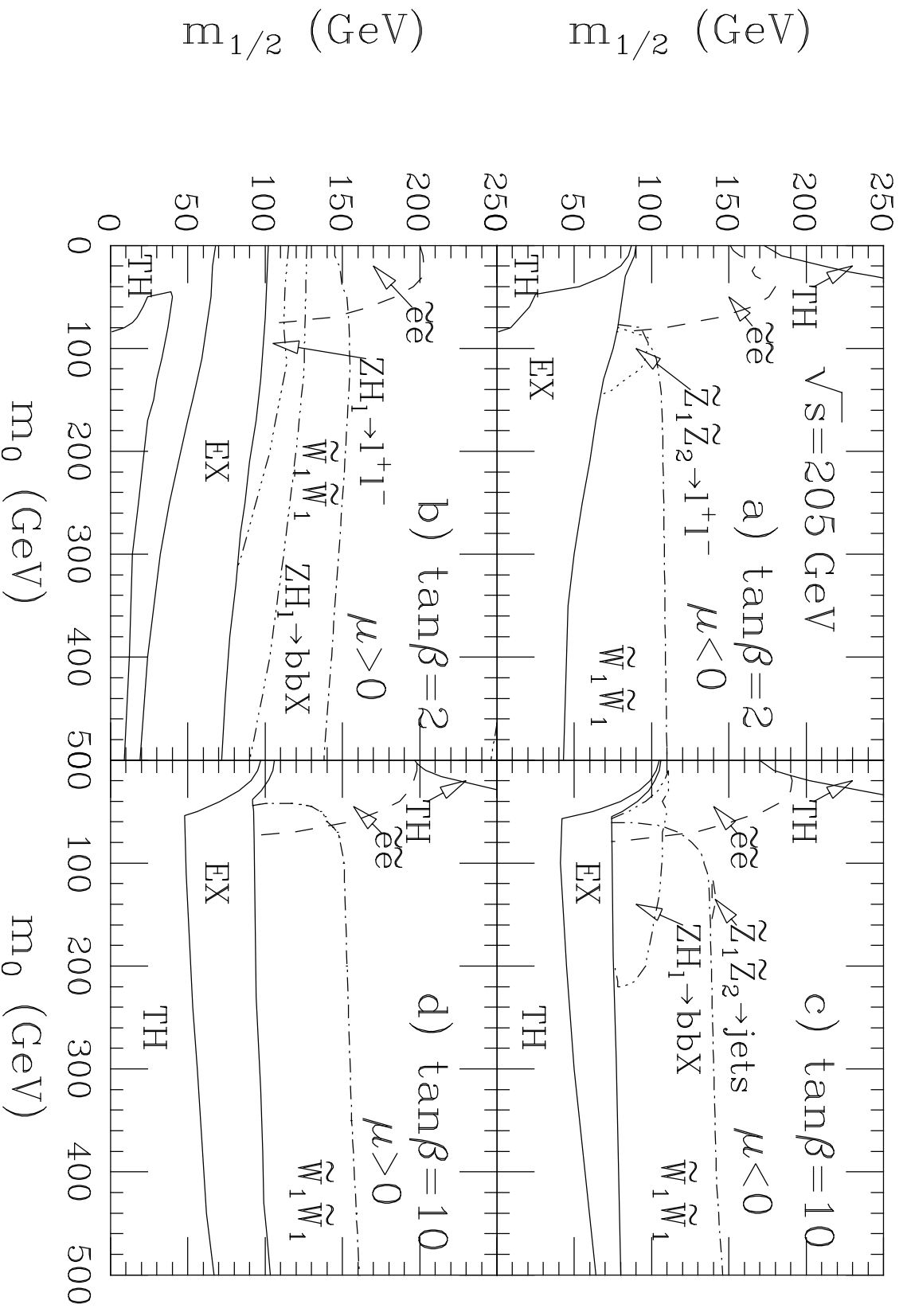
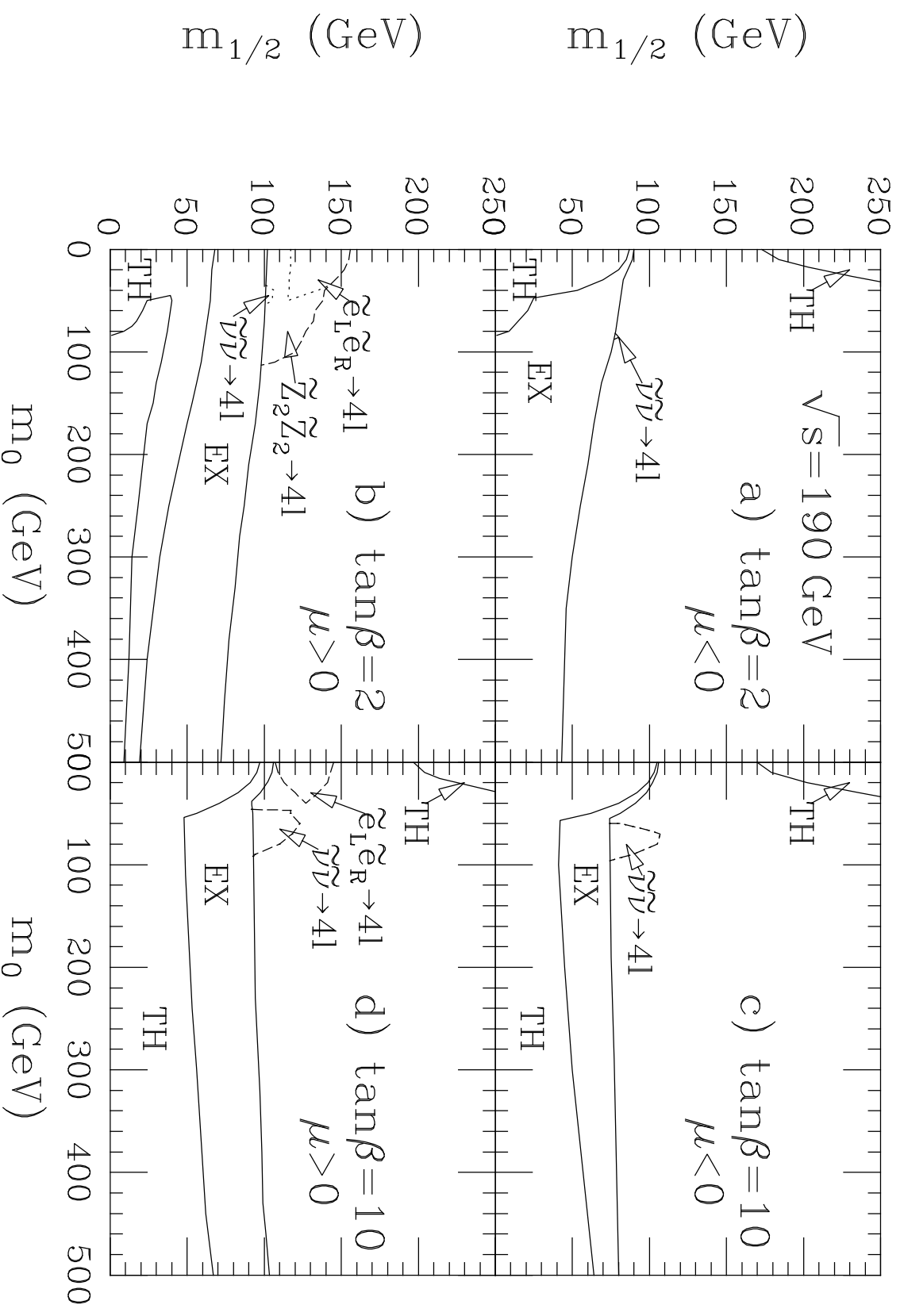
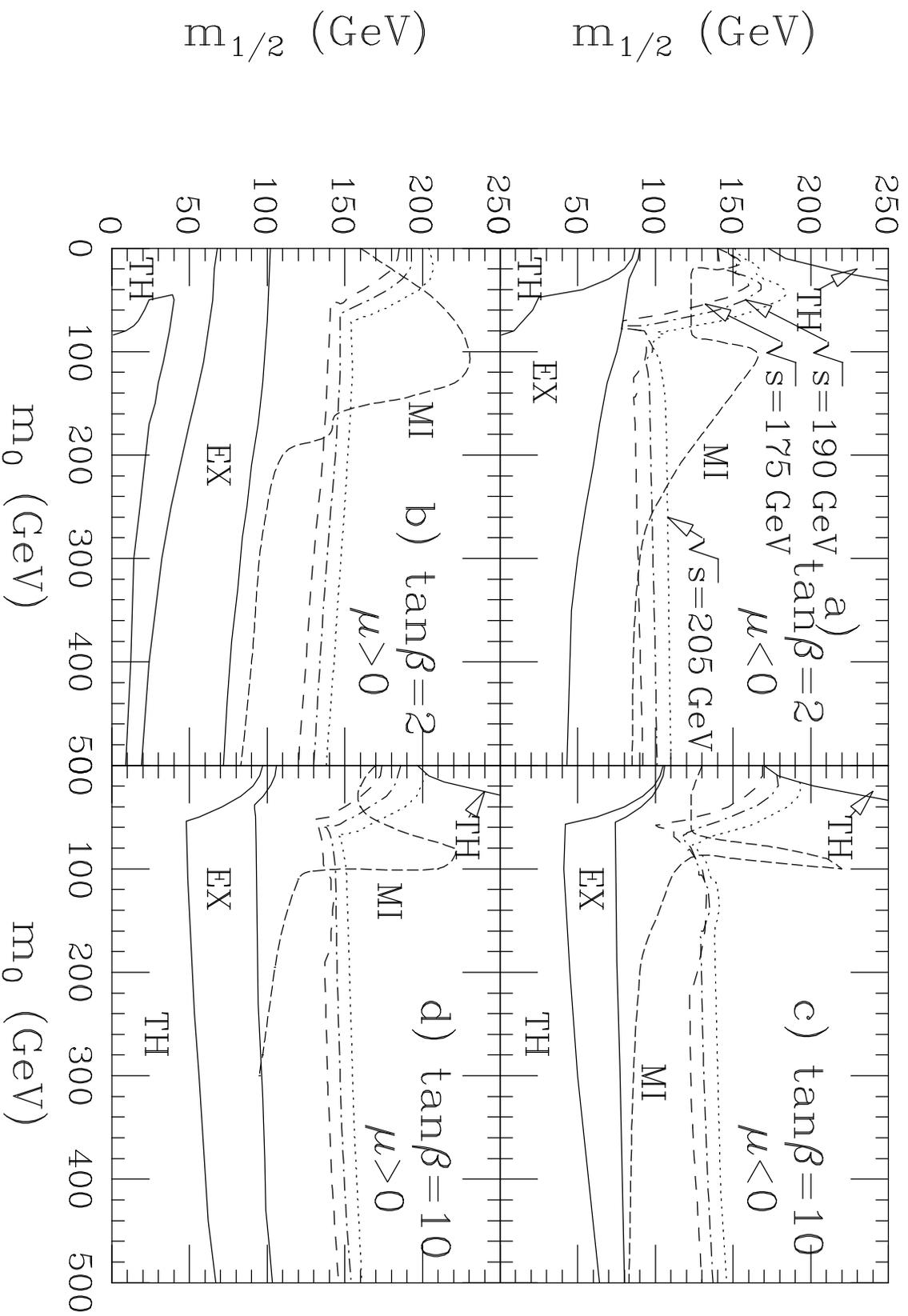
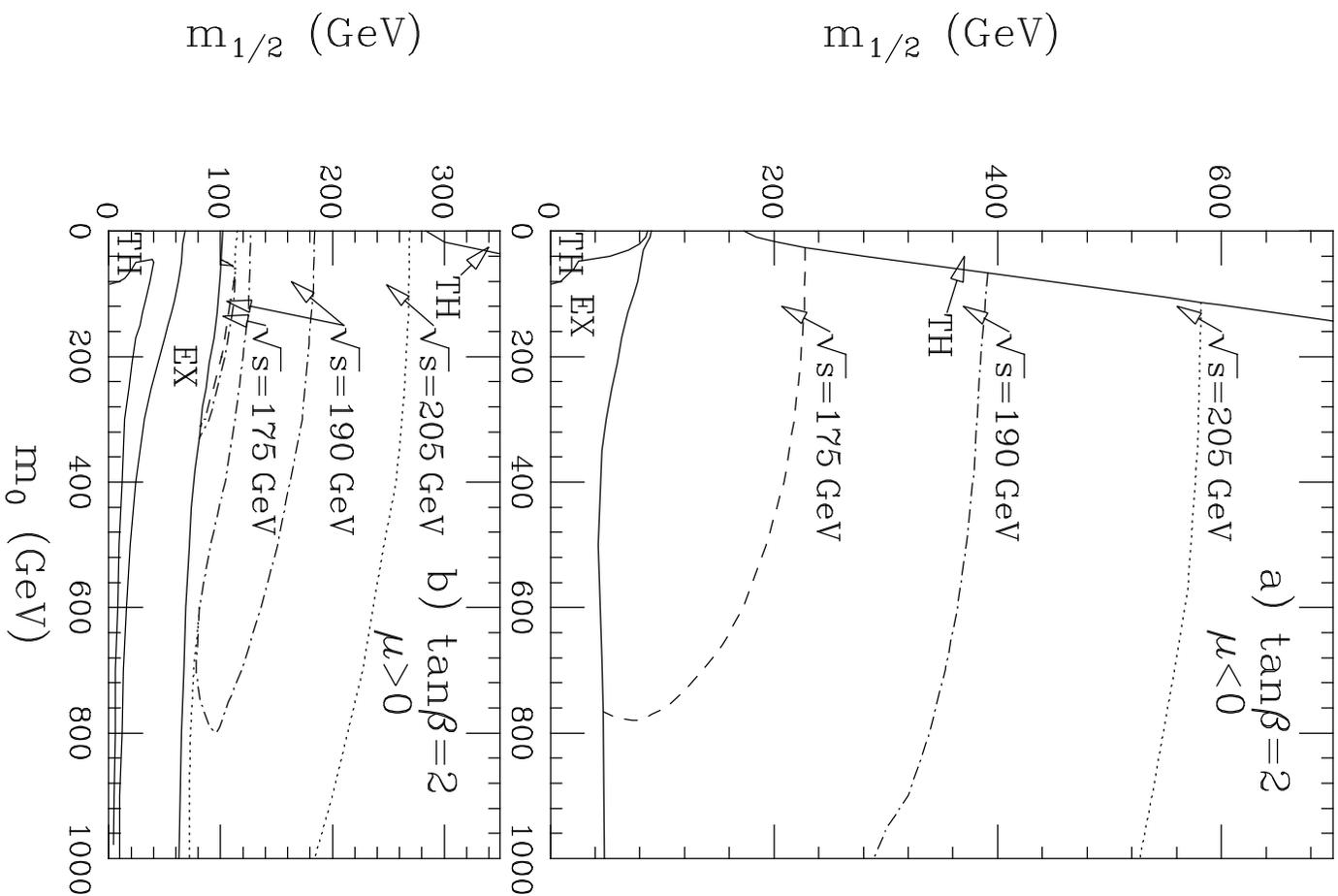
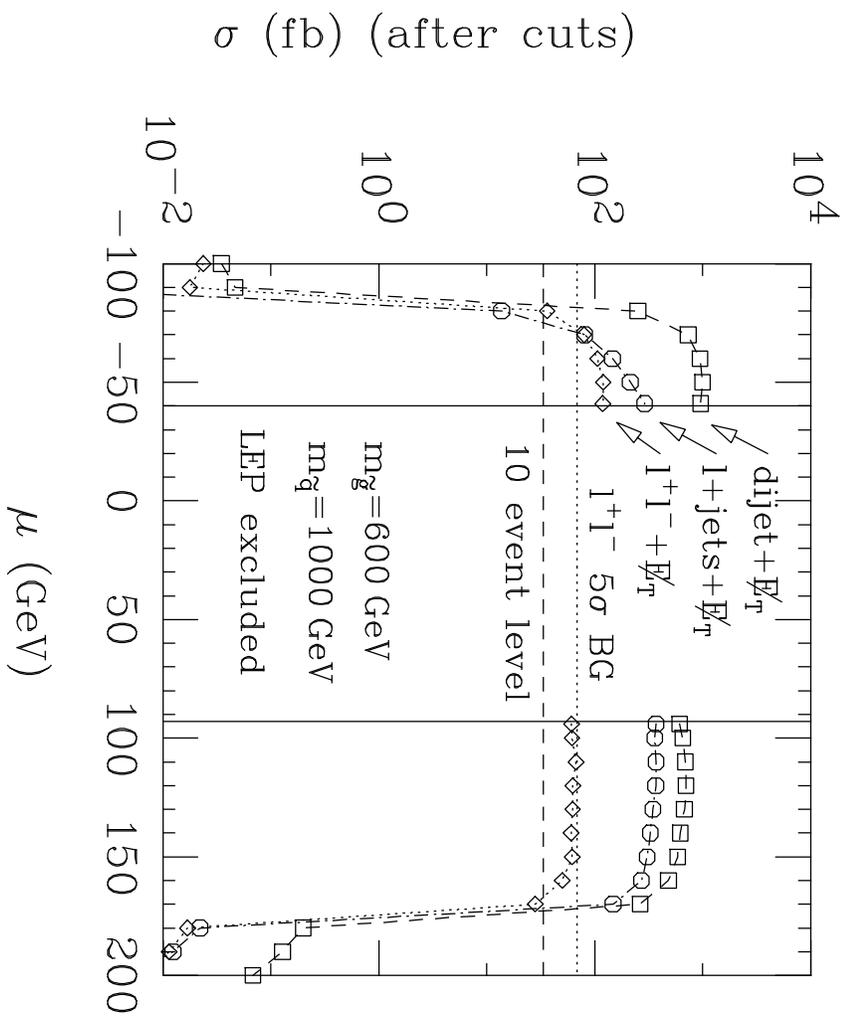
\end{document}